\documentclass[aps,prb,twocolumn,amsmath,nofootinbib,longbibliography,amssymb,superscriptaddress,10pt,floatfix]{revtex4-2}
\usepackage[colorlinks, linkcolor=Blue , citecolor= Blue,urlcolor = NavyBlue, breaklinks=true]{hyperref}
\usepackage[english]{babel}
\usepackage{ulem}
\usepackage[svgnames]{xcolor}
\usepackage{verbatim}
\usepackage{braket}
\usepackage{esint}
\usepackage{mathtools,bm,bbm}
\usepackage{multirow}

\usepackage{tikz-qtree}
\usepackage{tikz}
\usepackage[compat=1.0.0]{tikz-feynman}
\usetikzlibrary{shapes.geometric, arrows.meta, decorations.markings}

\tikzset{
  thinarrow/.style={
    decoration={
      markings,
      mark=at position 0.5 with {\arrow{Stealth[length=1.3mm, width=0.8mm]}}
    },
    postaction=decorate
  },
  thickarrow/.style={
    decoration={
      markings,
      mark=at position 0.5 with {\arrow{Stealth[length=2.4mm, width=1.6mm]}}
    },
    postaction=decorate
  },
  doubledashed/.style={
    double,
    double distance=1.2pt,
    dashed,
    thickarrow
  }
}

\begin{document}
\title{
Controlled Loop Expansion for Strained Twisted Bilayer Graphene
}
\author{Eyal Keshet}
\affiliation{Department of Condensed Matter Physics, Weizmann Institute of Science, Rehovot 76100, Israel}
\author{Yaar Vituri}
\affiliation{Department of Condensed Matter Physics, Weizmann Institute of Science, Rehovot 76100, Israel}
\author{Erez Berg}
\affiliation{Department of Condensed Matter Physics, Weizmann Institute of Science, Rehovot 76100, Israel}
\affiliation{Materials Department, University of California Santa Barbara, Santa Barbara 93106 USA}
\affiliation{Department of Electrical and Computer Engineering, University of California, Santa Barbara, CA 93106,
USA}

\date{\today}

\begin{abstract}
We develop a controlled diagrammatic framework for periodic Anderson models, 
and apply it to heterostrained magic-angle twisted bilayer graphene (MATBG) at charge neutrality using the topological heavy-fermion formulation. Building 
on \href{https://arxiv.org/abs/2604.14278}{arXiv:2604.14278}, 
we organize self-energy insertions and perform a Dyson resummation to any order in the small parameter $s^2$ -- the fraction of the moir\'e Brillouin zone with nontrivial quantum geometry. For strained MATBG, the expansion remains controlled down to arbitrarily low temperatures as long as the strain induced energy scale is not too small. In the flat-chiral limit, an emergent approximate $\rm{U}(1)$ symmetry forbids the leading scattering channel and leaves the Mott bands sharp at order $s^2$. This is in stark contrast to the unstrained case, where the linewidth is of order $N_f s^2 U$ with $U$ the on-site $f$-$f$ Hubbard interaction and $N_f$ the number of $f$ states per site. Away from the chiral limit, the linewidth is non-zero at order $s^2$ but more than an order of magnitude smaller than in the unstrained case. The strain-induced energy scale also imprints itself directly on the spectrum: as an electron-phonon-like kink in the dispersion, and as an additional flat ``trion'' band -- a single-particle excitation bound to a local $f$ particle-hole pair. We use the framework to predict the Quantum Twisting Microscope spectrum at one-loop order for both strained and unstrained MATBG, and compare with recent experiments. 
\end{abstract}

\maketitle

\section{Introduction}
The discovery of superconductivity and correlated insulators in magic-angle twisted bilayer graphene (MATBG)~\cite{cao2018correlated, cao2018unconventional} has launched a plethora of experiments exploring its unique phase diagram, which includes unconventional superconductivity~\cite{cao2018unconventional, yankowitz2019tuning, lu2019superconductors, codecido2019correlated, saito2020independent, cao2021nematicity, tian2023evidence, stepanov2020untying, arora2020superconductivity, liu2020tunable, park2022robust, cao2021pauli}, Chern insulators~\cite{grover2022chern, wu2021chern, xie2021fractional, pierce2021unconventional} and 
both integer and fractional quantum anomalous Hall
states~\cite{sharpe2019emergent, serlin2020intrinsic, stepanov2021competing}. In addition to being a system of strongly correlated electrons, MATBG is an attractive research candidate because it is highly tunable. Its carrier density is tunable using electrostatic gating, and alignment with a substrate has also been shown to affect its phase diagram~\cite{serlin2020intrinsic, sharpe2019emergent, stepanov2021competing, shi2021moire, wong2023insulators}. 

Among these tuning parameters, strain is unique. On the one hand, it has been shown to significantly change the phase diagram, stabilizing for example the incommensurate Kekul\'{e} spiral order~\cite{nuckolls2023quantum, kwan2021kekule, wagner2022global, kwan2024electron, herzog2025kekule, wang2025putting, wang2023ground}. On the other hand, unlike the density and substrate potential mentioned above, it is both difficult to control and ubiquitous in experiments, with different experimental samples exhibiting uniaxial heterostrain of 0\% to 0.7\%~\cite{kerelsky2019maximized, choi2019electronic, xie2019spectroscopic, oh2021evidence, wong2020cascade, xiao2025interacting, parker2021strain}. Thus, from a theoretical standpoint, studying the effects of strain is crucial for understanding experimental results.

Being a system of strongly correlated electrons, developing an analytically controlled description of MATBG remains challenging, with early theoretical progress largely restricted to zero-temperature~\cite{bultinck2020ground,kwan2021kekule,wagner2022global,shavit2021theory,kwan2024electron,xie2020nature, lian2021twisted}. An important step in this direction was the
topological heavy fermion (THF) reformulation of Song and Bernevig~\cite{song2022magic, cualuguaru2023twisted},
which recasts the low-energy bands of MATBG into localized $f$-electrons
hybridized with itinerant topological $c$-electrons.

A crucial further step towards an analytically controlled description of MATBG was taken by Ledwith
\textit{et al.}~\cite{ledwith2024nonlocal}, who identified a small parameter $s^2 \approx
5$-$10\%$ characterizing the fraction of the Brillouin zone with nontrivial quantum geometry of the low-energy bands. In the THF basis this parameter can be understood as the fraction of the Brillouin zone where the active bands' excitations are predominantly $c$-like. 
A recent paper by some of the present authors~\cite{vituri2026controlled}
built on this to develop a controlled diagrammatic expansion for the THF basis, borrowing elements from strong-coupling perturbation theory~\cite{pairault1998strong,pairault2000strong}.
The central object controlling the expansion is the hybridization integral $\mathcal I$, with $\rm{Im}[\mathcal I]$ the $c$-electron density of states weighted by the $c$-$f$ hybridization. The condition $|\mathcal{I}| \ll U$
defines the range of validity of the expansion, with $U$ the $f$-$f$ on-site interaction strength. In the flat-chiral limit of unstrained MATBG, it was shown that $\mathcal{I(\omega)} \sim
N_f s^2 \omega$ up to logarithmic corrections, and the small parameter that controls the expansion is $N_f s^2$ ($N_f$ being the number of $f$ states per site). Diagrammatically, the order in $s^2$ corresponds to the number of loops in a diagram. The expansion
was carried to one-loop order for unstrained MATBG in the limit of large $N_f$ by performing an infinite-loop resummation, yielding the quasiparticle
lifetime of the Mott bands~\cite{vituri2026controlled,wei2026lifetime, hu2026twisted, nosov2026controlled}. 

In this paper we extend this scheme in two directions. First, we present a
general and systematic scheme to perform the Dyson resummation of the $s^2$
expansion to any order, applicable to any periodic Anderson model with a small
hybridization phase space. This scheme is both simpler and more general than
the one used in Ref.~\cite{vituri2026controlled}, and does not require taking the limit of large $N_f$.
Second, we apply this framework to study strained MATBG at charge neutrality.

The strained case turns out to be different from the unstrained one in a number of interesting ways. In the flat-chiral limit with strain added, an
additional approximate $\rm{U}(1)$ symmetry severely restricts scattering, with striking
consequences for the quasiparticle lifetime that stand in sharp contrast to the
unstrained case. Away from this limit, new spectral features emerge that are
set by the strain energy scale, and which we analyze analytically. These spectral features include a kink in the dispersion reminiscent of the one seen in electron-phonon scattering, and for some parametric regimes, a second flat band that appears below the Hubbard band. This additional flat band is the result of a single-particle excitation mixing with a strain-induced trion --- a single-particle and exciton bound state. The exciton is a local $f$ particle-hole pair with an energy set by the strain energy scale. We thus dub this new flat band the ``trion" band.
For some model parameters, most of the spectral weight in the active bands away from the $\gamma_M$ point shifts to this new strain-induced trion band.

These spectral features are directly relevant to recent experiments. The
electronic spectral function of MATBG was recently measured
in a momentum- and energy-resolved measurement using a Quantum Twisting
Microscope (QTM)~\cite{inbar2023quantum,xiao2025interacting}. Given that real samples may carry strain of unknown
magnitude, a natural and pressing question is how important are the effects originating from strain to the observed spectrum. Using our framework we compute the predicted QTM spectrum for both
the strained and unstrained cases to order $s^2$ at charge neutrality, identify the key
qualitative differences between them, and compare to the experimental results.

The remainder of this paper is organized as follows. In
Sec.~\ref{sec:controlled_expansion} we develop the general $s^2$ expansion and Dyson
resummation scheme. In Sec.~\ref{sec:strained} we apply this to strained MATBG
at charge neutrality. In Sec.~\ref{sec:qtm} we present the predicted QTM
spectra for both the unstrained and strained cases and compare to experiment. We conclude with a discussion and outlook in Sec.~\ref{sec:discussion}. A summary of our main results is given in
Sec.~\ref{sec:summary}.

\section{Summary of Results}
\label{sec:summary}

Our starting point is a periodic Anderson model with $f$-$f$ on-site Hubbard interaction $U$. As in Ref.~\cite{vituri2026controlled}, we start by tracing out the $f$-electrons to get an effective theory for the $c$-electrons with an infinite number of frequency-dependent interaction vertices.

As shown in Ref.~\cite{vituri2026controlled}, in this effective $c$-only theory one can still calculate both $c$- and $f$-properties. Our goal in Sec.~\ref{sec:controlled_expansion} is to identify the diagrams in this effective theory that give the $f$-electrons' self-energy to any order in $s^2$ and non-perturbatively in $U$. Ref.~\cite{vituri2026controlled} started by calculating the self-energy of the $c$-electrons using 1PI diagrams in the effective $c$-only theory. In doing this, divergences are encountered at energy scales set by the $f$-electrons, and these were taken care of by an infinite loop resummation. Here, we relate these diagrams of the $c$-only theory to an infinite set of diagrams in the full $c$-$f$ theory with a fixed number of hybridization vertices and any number of $U$ insertions. We identify that these divergences are a result of contributions to the diagram that are \textit{1-particle-reducible} (1PR) with respect to the $f$-electrons. We develop a scheme of subtracting these 1PR contributions, obtaining the $f$ self-energy at any order in $s^2$ and performing a simple Dyson resummation to obtain the $f$ Green's function. As mentioned previously, the order in $s^2$ is equal to the number of $c$ loops in a given diagram. These results are summarized in Sec.~\ref{sec:controlled_expansion_C}. The $c$-electrons' self-energy and Green's function can then be obtained by simple relations already obtained in Ref.~\cite{vituri2026controlled}. We stress that the results of this section are general, and applicable to any periodic Anderson model where the hybridization integral is small.

In Sec.~\ref{sec:strained} we apply our results to strained MATBG at charge neutrality, using the THF formulation. In the THF language, $s^2$ takes the explicit form
\begin{equation}
    s^2 = \frac{\gamma^2}{2v_\star^2}\frac{2\pi}{A_\text{BZ}},
        \label{eq:s2_def}
\end{equation}
where $\gamma$ is the $f$-$c$ hybridization, $v_\star$ the Dirac
velocity of the $c$-electrons, and $A_\text{BZ}$ the moir\'e Brillouin zone area. The hybridization integral takes the form
\begin{equation}
    \mathcal{I}(i\omega) = \sum_\lambda \int \frac{d^2k}{A_\text{BZ}}
    \!\left[\hat\gamma_\mathbf{k}\,
    G_{c,\lambda}(\mathbf{k},i\omega)\,\hat\gamma^\dagger_\mathbf{k}\right],
    \label{eq:I_def}
\end{equation}
where $G_{c,\lambda}(\mathbf{k},i\omega)$ is the full $c$-electron propagator for
flavor $\lambda$, $\hat{\gamma}_\mathbf{k}$ is the hybridization
term between $f$ and $c$-electrons, and the sum runs over all flavors
$\lambda$. Note that Ref.~\cite{vituri2026controlled} defined $\mathcal I(i\omega)$ as the trace of the matrix defined here. For unstrained MATBG, this matrix is proportional to the identity matrix, therefore the condition $|\mathcal I|\ll U$ remains the same under the spectral norm. We study the model in the regime $U\gg M_f |\mathbf \epsilon|\gg Us^2, T$, where $|\mathbf \epsilon|$ is the magnitude of the heterostrain, $M_f$ a model parameter for the $f$ on-site strain-induced level splitting, and $T$ the temperature. In this regime, the anisotropic semimetal is the ground state, and the expansion remains valid to $T=0$.

We find that in the flat-chiral limit with only the on-site $f$-$f$ interaction MATBG possesses an additional $\rm{U}(1)$ symmetry. In the presence of strain, particle-hole excitations that are neutrally charged with respect to this symmetry have a gap of order $\sim\gamma$. As a result, the ground state becomes a Slater determinant in the projected limit, and the system's excitations are infinitely long-lived for finite $\gamma \gg U$. 

In Sec.~\ref{sec:strained_tree_level} we calculate the spectral function to tree-level, equivalent to the Hubbard-I approximation~\cite{hubbard1963electron}, as done in Ref.~\cite{ledwith2024nonlocal, ledwith2025exotic}. To order $s^0$, we find that the non-perturbative effect of the Hubbard interaction for single-particle excitations is the Hartree-Fock correction to the $f$ on-site energy. 
Consequently, the results to order $s^0$ are identical to the ones obtained using Hartree-Fock~\cite{Kang2020,liu2021nematic, xie2020nature, bultinck2020ground, parker2021strain, kwan2021kekule, herzog2025kekule}.

We follow this in Sec.~\ref{sec:strained_one_loop} and calculate the one-loop correction, focusing on long-range in imaginary time processes as done in Ref.~\cite{vituri2026controlled}. In the absence of strain, these are identified as zero-energy flavor-flips. The addition of strain adds an on-site $f$ single-particle term that splits the flat bands into low and high energy sectors. As a result, at charge neutrality there is no phase space for zero-energy flavor-flips, and the dominant long-range in time process becomes an orbital-flip 
with a non-zero energy cost 
of $2M_f|\epsilon|$. 
This results in striking contrast to the unstrained cases. For the strained case at the flat-chiral limit the Mott bands are sharp to order $s^2$, while away from that limit their width is proportional to the deviation from the flat-chiral limit. This is in contrast to the unstrained case, where the Mott bands' width is of order $N_f s^2 U$ in the flat-chiral limit~\cite{vituri2026controlled,wei2026lifetime, hu2026twisted, nosov2026controlled}. 

Furthermore, we find new spectral features in the strained case, analogous to those seen in the electron-phonon model with Einstein phonons~\cite{marsiglio2008electron}, with the orbital-flip mentioned previously acting as the constant-energy bosonic mode. In some parametric regimes we find a kink in the spectrum where the dispersion crosses the energy of the orbital-flip, while at other regimes an additional flat band appears. We call this new flat band a trion band, as it arises from single-particle excitations hybridizing with bound states of single-particles and particle-hole orbital-flips. The location of these features is set by the strain magnitude, and we explain their origin in Sec.~\ref{sec:strained_features}. We note that in some parametric regimes, most of the spectral weight away from the $\gamma_M$ point shifts from the Hubbard band to this lower energy strain-induced trion band.

Finally in Sec.~\ref{sec:qtm} we calculate the QTM spectrum of both the strained and unstrained cases. Recent QTM experiments have measured the momentum- and energy-resolved spectral function of MATBG~\cite{xiao2025interacting}. It is an open question how significant the strain in the experimental samples is to the observed spectrum. Comparing the strained and unstrained QTM spectra, we find significant differences. First, the breaking of $C_{3z}$ symmetry in the strained case results in the QTM measurement showing three different spectral lines, as already seen in calculations using the non-interacting BM model~\cite{wei2025dirac}. Furthermore, the extra spectral features theoretically predicted in the strained case are visible in the calculated QTM spectrum. We can use these extra features predicted in the strained case to place severe restrictions on the model parameters and strain magnitude when comparing the theoretical predictions to experiment.

\section{Controlled $s^2$ Expansion}
\label{sec:controlled_expansion}

The main goal of this section is to establish a simple way to calculate the self-energy of a periodic Anderson model to any order in the hybridization integral.

We consider a generic multi-orbital, multi-flavor periodic Anderson model, with the action
\begin{equation}
    S = S_c + S_f + S_{fc},
\end{equation}
where $f$ denotes the localized electrons, $c$ the conduction electrons, and
\begin{equation}
\begin{split}
S_c &= \int_0^\beta d\tau \sum_{\mathbf k,\lambda}
\bar{c}_{\mathbf k,\lambda,a}
\Big[(\partial_\tau - \mu)\delta_{aa'} + h^{(\eta)}_{\mathbf k,aa'}\Big]
c_{\mathbf k,\lambda,a'}, 
\\[6pt]
S_f &= \int_0^\beta d\tau \sum_{\mathbf{R}, \lambda}
\Big[ \bar{f}_{\mathbf{R},\lambda,b} \big((\partial_\tau - \mu)\delta_{bb'}+h^{(\eta)}_{\mathbf R, bb'}\big) f_{\mathbf{R},\lambda,b'} \Big] \\
&\qquad\qquad\qquad\qquad\qquad\qquad\qquad\qquad+ \frac{U}{2}\, \delta n_\mathbf{R}^2, 
\\[6pt]
S_{fc}
&= \int_0^\beta d\tau \sum_{\textbf{k}, \lambda}
\Big[
\Bigl(\hat\gamma_{\mathbf{k}}\Bigr)_{b a}
\Biggl(\sum_{\mathbf{R}}\frac{e^{i\mathbf{k}\cdot\mathbf{R}}}{\sqrt{N_s}}\,
\bar f_{\mathbf{R},\lambda,b}(\tau)\Biggr)
c_{\mathbf{k},\lambda,a}(\tau) \\
&\qquad\qquad\qquad\qquad\qquad\qquad\qquad\qquad\;+ \text{h.c.}
\Big].
\label{eq:original_action}
\end{split}
\end{equation}

Here $a, a', b, b'$ are the orbital indices of the $c$ and $f$-electrons, respectively, $\lambda=1,..., 4$ the flavors, $R$ denotes lattice positions, $\delta n_\mathbf{R}\equiv\sum_{\lambda, b}\left(\bar f_{\mathbf{R},\lambda,b} \,f_{\mathbf{R},\lambda,b}-\frac{1}{2}\right)$. The single-particle Hamiltonians $h^{(\eta)}_{\mathbf k,aa'}, h^{(\eta)}_{\mathbf R, bb'}$ are kept general for now.
We define $c_{\mathbf R,\lambda,a}$ such that
\begin{equation}
    c_{\mathbf k,\lambda,a} = 
    \sum_R \frac{e^{-i\textbf{k}\cdot\mathbf{R}}}{\sqrt{N_s}} c_{\mathbf R,\lambda,a}
\end{equation}
for any $\mathbf{k}$ within the first Brillouin zone. We further restrict all summations over momentum to the first Brillouin zone for simplicity such that $\{c_{\mathbf R,\lambda,a}\}$ spans all $c$-electron annihilation operators in the theory. This restriction is not essential.

\subsection{$c$-only Action}
\label{sec:controlled_expansion_A}

To obtain an effective action for the $c$-electrons, we follow the strong-coupling expansion~\cite{pairault1998strong, pairault2000strong, senechal2002cluster,Valenti1993} and trace out the local $f$-electrons~\cite{vituri2026controlled}. Here and throughout, $\langle\rangle_{0}$ denotes correlators in the unhybridized theory given by $\hat \gamma_{ab}=0$. $\langle\rangle_c$ and $\langle\rangle_{c,0}$ denote connected correlators in the full and unhybridized theories, respectively. Note that in the unhybridized theory the $f$ action is completely local, and therefore the connected correlators $\langle\rangle_{c,0}$ must also be local.

The result of integrating out the $f$-electrons is an effective action that contains an infinite number of vertices composed of local $f$ correlators. The action is given by: 
\begin{equation}
    S_{\rm{eff}} = S_c - \log\big\langle e^{-S_{fc}}\big\rangle_0,
\end{equation}
with
\begin{widetext}
\begin{equation}
\begin{split}
\log\big\langle e^{-S_{fc}}\big\rangle_0
= \sum_{R} \sum_{n=1}^{\infty} \frac{1}{(n!)^2}&
\sum_{\{i\},\{i'\}}
\int_0^\beta \prod_{j=1}^n d\tau_j\,d\tau'_j\; \times \\
&\times
\bar c_{R,\lambda_n,a_n}(\tau_n)\cdots \bar c_{R,\lambda_1,a_1}(\tau_1)\;
\Gamma^{(n)}_{1\cdots n;\,1'\cdots n'}\;
c_{R,\lambda'_1,a'_1}(\tau'_1)\cdots c_{R,\lambda'_n,a'_n}(\tau'_n)
\label{eq:c_only_action}
\end{split}
\end{equation}
and the $n$-body vertex

\begin{align}
\Gamma^{(n)}_{1\cdots n;\,1'\cdots n'}
&=
\sum_{\{b_j\},\{b'_j\}}
\prod_{j=1}^n (\hat \gamma^\dagger)_{a_j b_j}
\left\langle
f_{R,\lambda_1,b_1}(\tau_1)\cdots f_{R,\lambda_n,b_n}(\tau_n)\,
\bar f_{R,\lambda'_{n},b'_{n}}(\tau'_n)\cdots
\bar f_{R,\lambda'_{1},b'_{1}}(\tau'_1)
\right\rangle_{c,0}
\prod_{j=1}^n \hat \gamma_{b'_j a'_j}.
\label{eq:gamma_vertex}
\end{align}
\end{widetext}
We abbreviate $i=(\lambda_i, a_i)$ in the vertex notation. 

As was shown previously~\cite{vituri2026controlled}, this formulation is beneficial when the hybridization integral $\mathcal I$, defined in Eq.~\eqref{eq:I_def}, is small compared to the excitation energy of a single $f$ site (typically of order $U$).
In this case, we can perform a perturbative expansion, treating the quadratic terms $S_c$ and $\Gamma^{(1)}$ exactly and expanding in the number of loops. 
In the case of unstrained MATBG, Eq.~\eqref{eq:I_def} was shown to be proportional to $N_f s^2 \omega$, where $N_f$ is the number of $f$-flavors and $s^2$ the small parameter defined in \eqref{eq:s2_def}.\footnote{Strictly speaking this was shown for unstrained MATBG in the flat-chiral limit, but corrections away from the flat-chiral limit are expected to be small~\cite{vituri2026controlled}.} 

At tree-level, this expansion reproduces the Hubbard-I approximation~\cite{hubbard1963electron, vituri2026controlled}. Performing such a perturbative expansion to one-loop order, we encounter divergences when calculating the $c$-electrons' self-energy at the single-site excitation energies. This happens precisely because we integrated out a degree of freedom that can be excited at those energies. To solve this problem in the case of unstrained MATBG, Ref.~\cite{vituri2026controlled} identified an infinite series of diverging contributions and resummed it to obtain the self-energy to order $s^2$ in the limit of large $N_f$. Here present a general way to calculate the self-energy of the $c$- and $f$-electrons in the model \eqref{eq:original_action} to any order in $s^2$ and non-perturbatively in $U$. To do this, we relate the diagrams of the $c$-only theory \eqref{eq:c_only_action} to those of the full $c$-$f$ theory, where it is easier to identify the part of the self-energy that is 1PI with respect to $f$.  
The first step is to consider a perturbative expansion of the full $c$-$f$ theory.

\subsection{Perturbative Expansion from the Full $f$-$c$ Action}
\label{sec:controlled_expansion_B}

We begin by going back to the action \eqref{eq:original_action}, formally performing a double perturbative expansion in $\hat \gamma, U$, such that we get terms of all orders $(\hat\gamma \hat \gamma ^ \dagger)^m U^l$.

As we consider $\hat\gamma, U$ that are both large and non-perturbative, we perform the calculation to all orders in $U$, and for $\hat \gamma$ we must reorganize the resulting perturbation theory in powers of the small parameter  $s^2$, or in other words the hybridization integral corresponding to \eqref{eq:I_def}. For a given order $m$ in the hybridization, there are contributions in a variety of orders in $s^2$. Furthermore, at each order in $m$ or in $s^2$ there is an infinite number of diagrams we must sum in order to take $U$ contributions to infinite order. Therefore, we cannot use simple diagrammatics and Wick's theorem. By examining each infinite set of diagrams in a particular order in $s^2$, we obtain expressions involving various $f$ $n$-point correlation functions, establishing the link to the $c$-only theory presented in Sec. \ref{sec:controlled_expansion_A}. By evaluating these correlation functions exactly in the single-site problem, our calculations are made non-perturbative in $U$.

In the next section, we explain diagrammatically how to perform this reordering process.

\subsection{Systematic $s^2$ Expansion}
\label{sec:controlled_expansion_C}

We start with a few preliminary definitions and relations. As established in Ref.~\cite{vituri2026controlled}, the $c$-electrons' self-energy is related to that of the $f$-electrons by the relation:
\begin{equation}
    \Sigma_c(\mathbf{k}, \omega) = \hat \gamma^\dagger_\mathbf{k} D(\mathbf{k}, \omega)\hat \gamma_\mathbf{k}
\label{eq:c_self_energy}
\end{equation}
With $D$ the irreducible part of $G_f$ with respect to the bare $c$-propagator
\begin{equation}
    G_f = \frac{1}{D^{-1}-\hat \gamma G_{c,0}\hat \gamma ^ \dagger}.
    \label{eq:D_def}
\end{equation} 
Here, $G_{c,0}$ is the bare $c$-propagator, and we dropped the $\mathbf k, \omega$ dependence for brevity.
This is a direct result of the fact that we have no $c$ interactions. We define the $f$ self-energy as the irreducible part of $G_f$ with respect to the full single site $f$-propagator, meaning the bare $f$-propagator fully dressed by the $U$ interaction:
\begin{equation}
    G_f = \frac{1}{G_{f,0}^{-1}-\Sigma_f},
    \label{eq:f_propagator_with_self_energy}
\end{equation}

Where $G_{f,0}$ is the single site local propagator.
Together with \eqref{eq:D_def}, this gives 
\begin{equation}
    \Sigma_f=\hat \gamma G_{c,0}\hat \gamma ^ \dagger+G_{f,0}^{-1}-D^{-1}.
\label{eq:f_self_energy_with_D}
\end{equation}

\subsubsection{First order in the hybridization}

We look at different contributions to the full $f$-propagator and self-energy at different orders $m$ in the hybridization. At order $m=1$ in the hybridization we have a single $c$-propagator, and example diagrams at this order are given by:
\begin{equation}
\tikzfeynmanset{
  thinarrow/.style={
    /tikz/decoration={
      markings,
      mark=at position 0.5 with {\arrow{Stealth[length=1.3mm, width=0.8mm]}}
    },
    /tikz/postaction=decorate
  }
}
\centering
\begin{tikzpicture}[scale=0.8, transform shape, baseline=(current bounding box.center)]
  \begin{feynman}

    
    \node at (-1.2, 0) {$G_{f} =$};
\begin{scope}[xshift=-0.5cm]
    \vertex (a1) at (0, 0);
    \vertex (b1) at (0.9, 0);
    \vertex (c1) at (2.4, 0);
    \vertex (d1) at (3.3, 0);

    \diagram* {
      (a1) -- [thinarrow] (b1) -- [dashed, thinarrow] (c1) -- [thinarrow] (d1)
    };
    \node [fill=white, inner sep=-0.6pt] at (b1) {\footnotesize $\times$};
    \node [fill=white, inner sep=-0.6pt] at (c1) {\footnotesize $\times$};
    \node [below=3pt] at (b1) {\footnotesize $\gamma$};
    \node [below=3pt] at (c1) {\footnotesize $\hat \gamma$};

    \node at (3.8, 0) {$+$}; 

    \vertex (a2) at (4.3, 0); 
    \vertex (b2) at (4.8, 0);
    \vertex (c2) at (7.6, 0);
    \vertex (d2) at (8.1, 0);
    \vertex (top1) at (5.5, 0.6);
    \vertex (top2) at (6.9, 0.6);

    \diagram* {
        (a2) -- [thinarrow] (b2) -- [thinarrow] (c2) -- [thinarrow] (d2),
        (b2) -- [bend right=60, thinarrow] (c2), 
        (b2) -- [thinarrow] (top1) -- [dashed, thinarrow] (top2) -- [thinarrow] (c2)
    };
    \node [fill=white, inner sep=-0.6pt] at (top1) {\footnotesize $\times$};
    \node [fill=white, inner sep=-0.6pt] at (top2) {\footnotesize $\times$};
    \node [above=2pt] at (top1) {\footnotesize $\hat \gamma$};
    \node [above=2pt] at (top2) {\footnotesize $\hat \gamma$};
    \node [below=3pt] at (b2) {\footnotesize $U$};
    \node [below=3pt] at (c2) {\footnotesize $U$};

    \node at (8.6, 0) {$+$};

    
    \begin{scope}[yshift=-2.5cm]
    \vertex (a4) at (0, 0); 
    \vertex (b4) at (0.5, 0);
    \vertex (mid5) at (1.9, 0);
    \vertex (c4) at (3.3, 0);
    \vertex (d4) at (3.8, 0);
    \vertex (top4a) at (1.2, 0.6);
    \vertex (top4b) at (2.6, 0.6);

    \diagram* {
        (a4) -- [thinarrow] (b4) -- [thinarrow] (mid5) -- [thinarrow] (c4) -- [thinarrow] (d4),
        (b4) -- [out=-60, in = -120, thinarrow] (mid5) --[out=-60, in = -120, thinarrow] (c4), 
        (b4) -- [thinarrow] (top4a) -- [dashed, thinarrow] (top4b) -- [thinarrow] (c4)
    };
    \node [fill=white, inner sep=-0.6pt] at (top4a) {\footnotesize $\times$};
    \node [fill=white, inner sep=-0.6pt] at (top4b) {\footnotesize $\times$};
    \node [above=2pt] at (top4a) {\footnotesize $\hat \gamma$};
    \node [above=2pt] at (top4b) {\footnotesize $\hat \gamma$};
    \node [below=3pt] at (b4) {\footnotesize $U$};
    \node [below=3pt] at (c4) {\footnotesize $U$};
    \node [below=3pt] at (mid5) {\footnotesize $U$};
    \end{scope}

    \node at (4.1, -2.5) {$+$}; 

    \begin{scope}[yshift=-2.5cm, xshift=4.6cm]
    \vertex (a3) at (0, 0); 
    \vertex (b3) at (0.5, 0);
    \vertex (mid3) at (1.9, 0);
    \vertex (mid4) at (1.9,-0.6);
    \vertex (c3) at (3.3, 0);
    \vertex (d3) at (3.8, 0);
    \vertex (top3a) at (1.2, 0.6);
    \vertex (top3b) at (2.6, 0.6);

    \diagram* {
        (a3) -- [thinarrow] (b3) -- [thinarrow] (mid3) -- [thinarrow] (c3) -- [thinarrow] (d3),
        (b3) -- [thinarrow, in = 180, out = -60] (mid4) -- [thinarrow, out=0, in=-120] (c3), 
        (b3) -- [thinarrow] (top3a) -- [dashed, thinarrow] (top3b) -- [thinarrow] (c3),
        (mid3) -- [thinarrow, out = -110, in=110] (mid4) -- [thinarrow, out = 70, in = -70] (mid3)
    };
    \node [fill=white, inner sep=-0.6pt] at (top3a) {\footnotesize $\times$};
    \node [fill=white, inner sep=-0.6pt] at (top3b) {\footnotesize $\times$};
    \node [above=2pt] at (top3a) {\footnotesize $\hat \gamma$};
    \node [above=2pt] at (top3b) {\footnotesize $\hat \gamma$};
    \node [below=3pt] at (b3) {\footnotesize $U$};
    \node [below=3pt] at (c3) {\footnotesize $U$};
    \node [above=3pt] at (mid3) {\footnotesize $U$};
    \node [below=3pt] at (mid4) {\footnotesize $U$};
    \end{scope}

    \node at (8.9, -2.5) {$+ \dots$};

\end{scope}
  \end{feynman}
\end{tikzpicture}
\end{equation}
where the solid lines are bare $f$-propagators without the $U$ interaction, the dashed lines are bare $c$-propagators, and the $\hat \gamma,\,U$ vertices are denoted accordingly. There are an infinite number of additional diagrams with any number of $U$ insertions.

In the first diagram, the momentum of the $c$-propagator is fixed by the outside momentum, so this is a contribution of order $s^0$. As shown previously~\cite{hu2025projected, vituri2026controlled}, this contribution gives the Hubbard-I propagator. 
In the other diagrams the momentum of the $c$-propagator is integrated over, therefore these diagrams are of order $s^2$, corresponding to one-loop diagrams in the $c$-only theory. We can identify that the set of diagrams obtained by ``cutting'' the $c$ line is a subset of the diagrams that contribute to the connected $f$ 4-point correlation function in the single site problem. We conclude that in order to sum to infinite order in $U$ we need to use the connected $f$ $4$-point correlation function, and the contributions from $m=1$ are therefore given by 
\begin{equation}
\tikzfeynmanset{
  thinarrow/.style={
    /tikz/decoration={
      markings,
      mark=at position 0.5 with {\arrow{Stealth[length=1.3mm, width=0.8mm]}}
    },
    /tikz/postaction=decorate
  },
  thickarrow/.style={
    /tikz/decoration={
      markings,
      mark=at position 0.5 with {\arrow{Stealth[length=2.4mm, width=1.6mm]}}
    },
    /tikz/postaction=decorate
  },
  doubledashed/.style={
    /tikz/double,
    /tikz/double distance=1.2pt,
    /tikz/dashed,
    thickarrow
  }
}
\centering
\begin{tikzpicture}[scale=0.8, transform shape, baseline=(a1.base)]
  \begin{feynman}

    \node at (-0.9, 0) {$G_{f} =$};

    \vertex (a1) at (0, 0);
    \vertex (b1) at (0.9, 0);
    \vertex (c1) at (2.4, 0);
    \vertex (d1) at (3.3, 0);

    \diagram* {
      (a1) -- [thinarrow] (b1) 
           -- [doubledashed] (c1) 
           -- [thinarrow] (d1)
    };
    
    \node [fill=white, inner sep=-0.6pt] at (b1) {\footnotesize $\times$};
    \node [fill=white, inner sep=-0.6pt] at (c1) {\footnotesize $\times$};
    \node [below=3pt] at (b1) {\footnotesize $\hat \gamma$};
    \node [below=3pt] at (c1) {\footnotesize $\hat \gamma$};

    \node at (3.8, 0) {$+$}; 

    \begin{scope}[xshift=5.2cm]
      \node[draw, rectangle, rounded corners=1pt, fill=white,
            inner sep=2pt, minimum width=12mm, minimum height=10mm] (B) at (0,0) {$\Gamma^{(2)}$};

      \diagram*{
        (B.north west) -- [doubledashed, out=90, in=90, looseness=1.5] (B.north east),
      };
    \end{scope}

    \node at (6.8, 0) {$+\,\dots$};
\label{eq:m1_diagrams}
  \end{feynman}
\end{tikzpicture}
\end{equation}
where we used the definition of the $\Gamma^{(2)}$ vertex from the effective $c$-only theory. When we draw such a $\Gamma^{(n)}$ vertex with missing outgoing legs, we mean to strip the vertex of the relevant factor of $\hat \gamma$ or $\hat \gamma^\dagger$ corresponding to these legs. For example, in the diagram to the right in \eqref{eq:m1_diagrams}, $\Gamma^{(2)}$ gives a factor $\Gamma^{(2)}
=
(\hat \gamma^\dagger)
\left\langle
f(\omega)\,
f(\omega')\,
\bar f(\omega')\,
\bar f(\omega)
\right\rangle_{c,0}
\hat \gamma,$
where we dropped all indices for brevity. Note that the original $\Gamma^{(2)}$ included four factors of $\hat \gamma$, and here we are left with two. The double dashed line denotes the full $c$-propagator $G_c$, so these are skeleton diagrams.

These diagrams give contributions to the full propagator, and to calculate the self-energy, we must drop the outside propagators for the $s^0$ contribution. For the $\Gamma^{(2)}$ contribution we cannot do that, and so we must multiply it by $G_{f,0}^{-2}$ to cancel the contribution of the incoming and outgoing propagators. This insight explains the double pole divergence seen in the calculation of the one-loop correction $D^{(1)}$ in Ref.~\cite{vituri2026controlled} as coming from the incoming and outgoing $f$-propagators. Overall, the contributions to the self-energy are then:

\begin{equation}
\tikzfeynmanset{
  thinarrow/.style={
    /tikz/decoration={
      markings,
      mark=at position 0.5 with {\arrow{Stealth[length=1.3mm, width=0.8mm]}}
    },
    /tikz/postaction=decorate
  },
  thickarrow/.style={
    /tikz/decoration={
      markings,
      mark=at position 0.5 with {\arrow{Stealth[length=2.4mm, width=1.6mm]}}
    },
    /tikz/postaction=decorate
  },
  doubledashed/.style={
    /tikz/double,
    /tikz/double distance=1.2pt,
    /tikz/dashed,
    thickarrow
  }
}
\centering
\begin{tikzpicture}[scale=0.8, transform shape, baseline=(b1.base)]
  \begin{feynman}

    \node at (-0.2, 0) {$\Sigma_{f} =$};

    \vertex (b1) at (0.6, 0);
    \vertex (c1) at (2.1, 0);

    \diagram* {
      (b1) -- [doubledashed] (c1)
    };
    
    \node [fill=white, inner sep=-0.6pt] at (b1) {\footnotesize $\times$};
    \node [fill=white, inner sep=-0.6pt] at (c1) {\footnotesize $\times$};
    \node [below=3pt] at (b1) {\footnotesize $\hat \gamma$};
    \node [below=3pt] at (c1) {\footnotesize $\hat \gamma$};

    \node at (2.8, 0) {$+$}; 
    \node at (4.0, 0) {$G_{f,0}^{-1}$};

    \begin{scope}[xshift=5.1cm]
      \node[draw, rectangle, rounded corners=1pt, fill=white,
            inner sep=2pt, minimum width=12mm, minimum height=10mm] (B) at (0,0) {$\Gamma^{(2)}$};

      \diagram*{
        (B.north west) -- [doubledashed, out=90, in=90, looseness=1.5] (B.north east),
      };
    \end{scope}

    \node at (6.2, 0) {$G_{f,0}^{-1}$};
    \node at (7.5, 0) {$+ \,\dots$};

  \end{feynman}
\end{tikzpicture}
\end{equation}

\subsubsection{Second order in the hybridization}

An example perturbative diagram at second order in the hybridization is given by:
\begin{equation}
\tikzfeynmanset{
  thinarrow/.style={
    /tikz/decoration={
      markings,
      mark=at position 0.5 with {\arrow{Stealth[length=1.3mm, width=0.8mm]}}
    },
    /tikz/postaction=decorate
  }
}
\centering
\begin{tikzpicture}[scale=0.8, transform shape, baseline=(a2.base)]
  \begin{feynman}

    \vertex (a2) at (0, 0);
    \vertex (b2) at (0.7, 0);     
    \vertex (mid1) at (1.7, 0);   
    \vertex (mid2) at (3.1, 0);   
    \vertex (c2) at (4.1, 0);     
    \vertex (d2) at (4.8, 0);

    \vertex (top1) at (1.7, 0.6);
    \vertex (top2) at (3.1, 0.6);

    \diagram* {
        (a2) -- [thinarrow] (b2) -- [thinarrow] (mid1) 
             -- [dashed, thinarrow] (mid2) 
             -- [thinarrow] (c2) -- [thinarrow] (d2),
        (b2) -- [bend right=50, thinarrow] (c2), 
        (b2) -- [thinarrow] (top1) 
             -- [dashed, thinarrow] (top2) 
             -- [thinarrow] (c2)
    };

    \node [fill=white, inner sep=-0.6pt] at (top1) {\footnotesize $\times$};
    \node [fill=white, inner sep=-0.6pt] at (top2) {\footnotesize $\times$};
    \node [fill=white, inner sep=-0.6pt] at (mid1) {\footnotesize $\times$};
    \node [fill=white, inner sep=-0.6pt] at (mid2) {\footnotesize $\times$};

    \node [above=2pt] at (top1) {\footnotesize $\hat \gamma$};
    \node [above=2pt] at (top2) {\footnotesize $\hat \gamma$};
    \node [above=2pt] at (mid1) {\footnotesize $\hat \gamma$};
    \node [above=2pt] at (mid2) {\footnotesize $\hat \gamma$};

    \node [below=4pt] at (b2) {\footnotesize $U$};
    \node [below=4pt] at (c2) {\footnotesize $U$};


  \end{feynman}
\end{tikzpicture}
\end{equation}

Similar to what was shown for the first order in the hybridization, this diagram is only one of many perturbative diagrams once we include all orders in $U$.  It is tempting to simply state that all such diagrams that scale as $s^4$ are given by the $\Gamma^{(3)}$ diagram
\begin{equation}
\centering
\begin{tikzpicture}[scale=0.8, transform shape, baseline=(H.base)]
  \begin{feynman}
    \node[draw,
          regular polygon,
          regular polygon sides=6,
          minimum width=15mm,
          minimum height=13mm,
          inner sep=3pt,
          fill=white] (H) at (0,0) {$\Gamma^{(3)}$};

    \draw[doubledashed] (H.corner 3) to[out=-190, in=-235, looseness=2.5] (H.corner 2);

    \draw[doubledashed] (H.corner 1) to[out=55, in=10, looseness=2.5] (H.corner 6);

\label{eq:mickey_mouse}
  \end{feynman}
\end{tikzpicture}
\end{equation}
While all the contributions from this diagram appear at order $s^4$ in the full $f$-propagator, the diagram \eqref{eq:mickey_mouse} also includes \textit{1-particle reducible} (1PR) diagrams with respect to $f$, such as
\begin{equation}
\centering
\begin{tikzpicture}[scale=0.7, transform shape, baseline=(a2.base)]
  \begin{feynman}
    \vertex (a2) at (0, 0);
    \vertex (b2) at (0.6, 0);      
    \vertex (mid1) at (1.6, 0);    
    \vertex (mid2) at (3.0, 0);    
    \vertex (c2) at (4.0, 0);      
    \vertex (top1) at (1.6, 0.6);  
    \vertex (top2) at (3.0, 0.6);  

    \vertex (conn) at (4.8, 0);

    \vertex (b2p) at (5.6, 0);     
    \vertex (mid1p) at (6.6, 0);   
    \vertex (mid2p) at (8.0, 0);   
    \vertex (c2p) at (9.0, 0);     
    \vertex (top1p) at (6.6, 0.6); 
    \vertex (top2p) at (8.0, 0.6); 
    \vertex (d2p) at (9.6, 0);

    \diagram* {
        (a2) -- [thinarrow] (b2) -- [thinarrow] (c2) 
             -- [thinarrow] (b2p)  
             -- [thinarrow] (c2p) -- [thinarrow] (d2p),
        
        (b2) -- [bend right=50, thinarrow] (c2),
        (b2) -- [thinarrow] (top1) -- [dashed, thinarrow] (top2) -- [thinarrow] (c2),

        (b2p) -- [bend right=50, thinarrow] (c2p),
        (b2p) -- [thinarrow] (top1p) -- [dashed, thinarrow] (top2p) -- [thinarrow] (c2p)
    };

    \foreach \n in {top1, top2, top1p, top2p} {
        \node [fill=white, inner sep=-0.6pt] at (\n) {\footnotesize $\times$};
    }

    \node [above=2pt] at (top1) {\footnotesize $\hat \gamma$};
    \node [above=2pt] at (top2) {\footnotesize $\hat \gamma$};
    \node [above=2pt] at (top1p) {\footnotesize $\hat \gamma$};
    \node [above=2pt] at (top2p) {\footnotesize $\hat \gamma$};
    

    \node [below=4pt] at (b2) {\footnotesize $U$};
    \node [below=4pt] at (c2) {\footnotesize $U$};
    \node [below=4pt] at (b2p) {\footnotesize $U$};
    \node [below=4pt] at (c2p) {\footnotesize $U$};

  \end{feynman}
\end{tikzpicture}
\end{equation}
This is a 1PR perturbative diagram that contains two 1PI ``sub-diagrams" and is part of \eqref{eq:mickey_mouse}, and it is one of the divergent diagrams that entered the resummation in Ref.~\cite{vituri2026controlled}. To get the 1PI contribution we must subtract the 1PR part, so the contribution to the self-energy can then be schematically written as:
\begin{equation}
\centering
\begin{tikzpicture}[scale=0.8, transform shape, baseline=(H.base)]
  \begin{feynman}
    \node at (-2.5, 0) {$\Sigma_f=\cdots +G_{f,0}^{-1}$};

    \node[draw,
          regular polygon,
          regular polygon sides=6,
          minimum width=15mm,
          minimum height=13mm,
          inner sep=3pt,
          fill=white] (H) at (0,0) {$\Gamma^{(3)}$};

    \draw[doubledashed] (H.corner 3) to[out=-190, in=-235, looseness=2.5] (H.corner 2);
    \draw[doubledashed] (H.corner 1) to[out=55, in=10, looseness=2.5] (H.corner 6);

    \node at (1.6, 0) {$G_{f,0}^{-1}$};

    \node at (2.6, 0) {$-\, G_{f,0}^{-1}$};

    \node at (3.3, 0) {$\Bigl($};

    \node[draw, rectangle, rounded corners=1pt, fill=white,
          inner sep=2pt, minimum width=12mm, minimum height=10mm] (B) at (4.2, 0) {$\Gamma^{(2)}$};

    \draw[doubledashed] (B.north west) to[out=90, in=90, looseness=1.5] (B.north east);

    \node at (5.3, 0) {$G_{f,0}^{-1}$};
    \node at (6.3, 0) {$\Bigr)^{2}+\cdots$};


  \end{feynman}
\end{tikzpicture}
\end{equation}
We need to use similar subtractions to obtain the 1PI contributions at any order in $s^2$.

\subsubsection{General reordering in powers of $s^2$}

We now generalize to all orders in $s^2$. The $m$-th order in the hybridization has contributions up to order $s^{2m}$. For a given diagram in the $c$-only theory, the order in $s^2$ is given by the number of independent loops in the diagram, and the number of $c$-propagators gives the order in the perturbative expansion in the hybridization where the corresponding contributions appear. To calculate the self-energy at some order, we draw skeleton diagrams in the effective $c$-only theory and use the full $c$-propagator, using the relation \eqref{eq:c_self_energy} to obtain it. 

As seen above, the $f$ $n$-point correlation functions contain parts that are 1PR with respect to $f$, which we need to subtract. The general rule is that at order $s^{2m}$ we need to subtract all the different ways one can order $k$ 1PI sub-diagrams of orders $n_1, n_2,...,n_k$ such that $\sum_{i=1}^kn_i=m$. Each subtraction comes with a combinatorial factor that gives the number of different ways one can order this set of sub-diagrams in a sequence, and an extra factor of $(-1)^{k+1}$ is needed to account for the fact that the subtractions happen recursively at all orders. Examples of such diagrammatic subtractions are given in App.~\ref{app:diagrams}. 

If we define
\begin{equation}
S=G_{f,0}^{-1}DG_{f,0}^{-1} - G_{f,0}^{-1},
\label{eq:D_S_relation}
\end{equation}
then $S$ is diagrammatically given by:
\begin{equation}
\centering
\begin{tikzpicture}[scale=0.8, transform shape, baseline=(R1.base)]
  \begin{feynman}
    \node at (-1.5, 0) {$G_{f,0}SG_{f,0} \;=\;$};

    \node[draw, rectangle, rounded corners=1pt, fill=white,
          inner sep=2pt, minimum width=11mm, minimum height=9mm] (R1) at (0.3, 0) {$\Gamma^{(2)}$};
    \draw[doubledashed] (R1.north west) to[out=90, in=90, looseness=1.5] (R1.north east);

    \node at (1.3, 0) {$+\,$};

    \node[draw, regular polygon, regular polygon sides=6, fill=white,
          minimum width=13mm, minimum height=11mm, inner sep=2pt] (H2) at (2.7, 0) {$\Gamma^{(3)}$};
    \draw[doubledashed] (H2.corner 3) to[out=-190, in=-235, looseness=2.5] (H2.corner 2);
    \draw[doubledashed] (H2.corner 1) to[out=55, in=10, looseness=2.5] (H2.corner 6);

    \node at (4.3, 0) {$+\,$};

    \begin{scope}[yshift=-2.4cm, xshift=0.5cm]
      \node[draw, rectangle, rounded corners=1pt, fill=white,
            inner sep=2pt, minimum width=11mm, minimum height=9mm] (EA) at (0, 0) {$\Gamma^{(2)}$};
      \node[draw, rectangle, rounded corners=1pt, fill=white,
            inner sep=2pt, minimum width=11mm, minimum height=9mm] (EB) at (2.0, 0) {$\Gamma^{(2)}$};

      \draw[doubledashed] (EA.north east) -- (EB.north west);
      \draw[doubledashed] (EA.south east) -- (EB.south west);
      \draw[doubledashed] (EA.north west) to[out=50, in=130] (EB.north east);

      \node at (3.1, 0) {$+\,$};

      \node[draw, regular polygon, regular polygon sides=8, fill=white,
            minimum width=15mm, minimum height=13mm, inner sep=2pt] (O3) at (4.8, 0) {$\Gamma^{(4)}$};
      \draw[doubledashed] (O3.corner 4) to[out=-170, in=-190, looseness=2.5] (O3.corner 3);
      \draw[doubledashed] (O3.corner 2) to[out=100, in=80, looseness=2.5] (O3.corner 1);
      \draw[doubledashed] (O3.corner 8) to[out=10, in=-10, looseness=2.5] (O3.corner 7);

      \node at (6.6, 0) {$+\;\cdots$};
    \end{scope}
\label{eq:S_definition}
  \end{feynman}
\end{tikzpicture}
\end{equation}
The self-energy series, including the combinatorial factors explained above, can be written as
\begin{equation}
    S - SG_{f,0}S+SG_{f,0}SG_{f,0}S+...=\frac{S}{1+G_{f,0}S}
    \label{eq:S_def_textual}
\end{equation}

Eq.~\eqref{eq:S_definition} gives the contribution to the self-energy beyond Hubbard-I, and thus the full $f$ self-energy is given by:
\begin{equation}
    \Sigma_f = \gamma_\mathbf k G_{c}\gamma_\mathbf k^\dagger + \frac{S}{1+G_{f,0}S}.
\label{eq:f_self_energy}
\end{equation}
At infinite order, both \eqref{eq:f_self_energy_with_D} and \eqref{eq:f_self_energy} give the same result for $\Sigma_f$. However, when truncating the series at finite order in $s^2$, expanding \eqref{eq:f_self_energy} takes care of the 1PR subtractions correctly.

To conclude, in order to calculate the self-energy, one can use the action \eqref{eq:c_only_action} and diagrammatically compute \eqref{eq:S_definition}. Inserting the result in \eqref{eq:f_self_energy} and using \eqref{eq:c_self_energy}, \eqref{eq:D_S_relation}, one obtains $\Sigma_f, \Sigma_c$

Finally, we note that a similar expression to \eqref{eq:f_self_energy} was obtained in a different context as part of the Dual Fermion Approach relating to DMFT~\cite{rubtsov2009dual, rohringer2018diagrammatic}.

\begin{figure*}
    \centering
    \includegraphics[width=\linewidth]{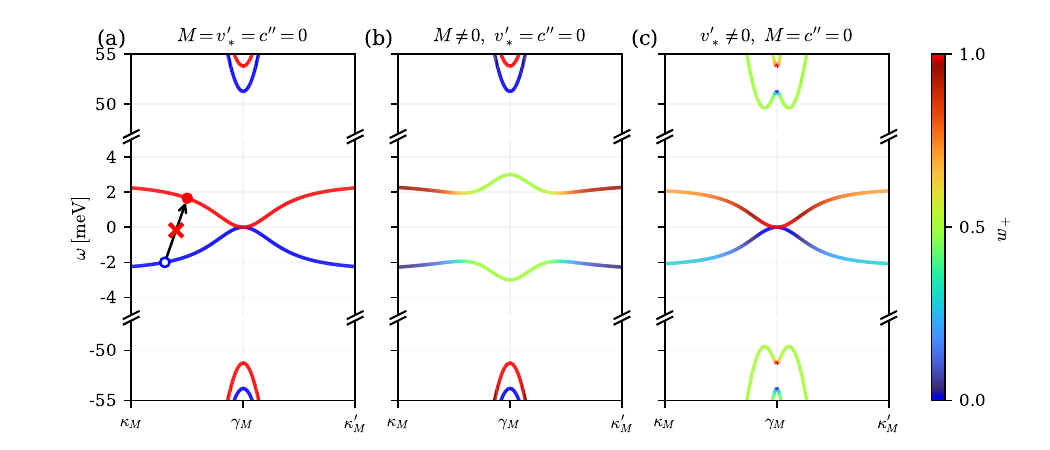}
    \caption{
    Spectral function $\mathcal{A}(\mathbf k,\omega)$ of the single-particle Hamiltonian 
    \eqref{eq:thf_hamiltonian}, obtained by diagonalizing the Hamiltonian. The colors represent the relative weight in the $\pm$ sectors of the symmetry defined in \eqref{eq:projections}, with red the $+$ sector, blue the $-$ sector, and the color scale in between represents an excitation with a mixed character. The panels show different parametric regimes, (a) the flat-chiral limit, (b) the chiral non-flat limit $M\ne0$, (c) the flat non-chiral limit $v_\star'\ne0$.
    In the flat-chiral limit the conduction and valence bands are in different sectors, and excitations such as the one shown by the black arrow are disallowed by the symmetry. $M\ne0$ results in sector mixing close to the $\gamma_M$ point, while $v_\star'$ results in mixing away from the $\gamma_M$ point. In (b) the Dirac points are off the cut shown here. When non-zero, all parameter values are taken from Table \ref{table:params}, with $|\epsilon|=0.1\%$ and $\epsilon_{xy}=0$.}
    \label{fig:ideal}
\end{figure*}

\section{Application to Strained MATBG at Charge Neutrality}
\label{sec:strained}
\subsection{Hamiltonian}
\label{sec:strained_ham}
We apply the results of the previous section to calculate the spectrum of strained MATBG. For this purpose, we use the Topological Heavy Fermion (THF) basis introduced and shown to faithfully represent MATBG by Song and Bernevig~\cite{song2022magic}. To add the effects of strain we use the minimal addition of parameters as shown in Ref.~\cite{herzog2025topological}, such that the entire model is given by

\begin{equation}
\label{eq:thf_hamiltonian}
\small 
\setlength{\arraycolsep}{2pt} 
\begin{aligned}
h^{(\eta)}_{\mathbf k} = 
\begin{pmatrix}
0_{2\times 2} & v_\star \sigma_x (\mathbf k^{(\eta)} \cdot \boldsymbol\sigma) & \gamma \sigma_0 + v_\star'(\mathbf k^{(\eta)} \cdot \boldsymbol\sigma) \\
v_\star (\mathbf k^{(\eta)} \cdot \boldsymbol\sigma)\sigma_x & M\sigma_x & c'' (\epsilon^{(\eta)} \cdot\boldsymbol\sigma)\sigma_x \\
\gamma \sigma_0 + v_\star'(\mathbf k^{(\eta)} \cdot \boldsymbol\sigma) & c''\sigma_x (\epsilon^{(\eta)} \cdot\boldsymbol\sigma) & M_f(\mathbf \epsilon^{(\eta)}\cdot\boldsymbol\sigma)
\end{pmatrix}
\end{aligned}
\end{equation}

Here, $\mathbf k^{(\eta)}=(\eta_z k_x, k_y)$ with $\eta_z$ being the valley index, $\mathbf \epsilon^{(\eta)}=(\epsilon_{xy}, \eta_z\epsilon_-)$, with $\epsilon_{ij}$ the strain tensor and $\epsilon_-=\frac{1}{2}(\epsilon_{xx}-\epsilon_{yy})$. The Hamiltonian is written in the basis where the first four entries correspond to the $c$ degrees of freedom, and the last two entries correspond to the $f$-electrons. $M_f, c''$ are the minimal set of parameters added to reproduce the strained spectrum of the Bistrizer-MacDonald (BM) model~\cite{herzog2025topological,bistritzer2011moire}. The Hamiltonian parameters used throughout this work are summarized in Table \ref{table:params}. 

A simplified version of this Hamiltonian is obtained in the strained flat-chiral limit\footnote{In this limit, in the presence of finite strain, the dispersion is in fact not flat. The word flat refers to the flatness in the zero strain limit. The chiral symmetry in this limit is present under finite strain.}, by setting $v_\star'=M=c''=0$. In the unstrained limit, this reduces to the flat-chiral limit of the THF basis as defined in Ref.~\cite{song2022magic}, and $c''=0$ is forced in the strained case by the chiral symmetry.

The full THF basis contains additional interaction terms, for example $V n_c n_c, W n_f n_c$, with $n_f, n_c$ the $f, c$ electron densities, respectively, as well as a flavor-exchange interaction $J$~\cite{song2022magic}. As shown in App.~\ref{app:interactions}, up to order $N_f s^2$ these give Hartree corrections that vanish at charge neutrality. There are additional non-Hartree corrections from the $W, J$ terms that renormalize the $c$-propagator and are relevant close to the $\gamma_M$ point but irrelevant to order $N_f s^2$ in the Mott bands. This contribution was calculated in Ref.~\cite{hu2026twisted}, and we neglect it here.

\begin{table*}[htbp]\centering
\begin{tabular}{c|c|c|c|c|c|c}
$v_\star$ & $M$ & $\gamma$ & $v_\star'$ & $U$ & $M_f$ & $c''$ \\
$-490\,[-2500]\,\mathrm{meV\cdot nm}$ & 
$2.8465\,\mathrm{meV}$ & 
$-52.5\,\mathrm{meV}$ & 
$158.0\,\mathrm{meV\cdot nm}$ & 
$30\,\mathrm{meV}$ & 
$4380\,\mathrm{meV}$ & 
$-3362\,\mathrm{meV}$ 
\end{tabular}
\caption{Hamiltonian parameter values. For $v_\star$, the value in brackets corresponds to the unstrained case, while the first value is for the strained case. Parameters are taken from Ref.~\cite{cualuguaru2023twisted} and to match the experimental results of Ref.~\cite{xiao2025interacting}. The unstrained $v_\star$ was scaled to decrease $s^2$ sufficiently, as done previously~\cite{hu2026twisted, vituri2026controlled}}
\label{table:params}
\end{table*}

\subsection{Additional Symmetry in the Flat-Chiral Limit}
\label{sec:strained_symmetry}
Consider the strained flat-chiral limit, in which the single-body Hamiltonian is given by

\begin{equation}
H(\mathbf{k})=
\begin{pmatrix}
0_{2\times 2} & v\,\sigma_x\!\big(\mathbf{k}^{(\lambda)}\!\cdot\!\boldsymbol{\sigma}\big) & \gamma\,\sigma_0 \\
v\,\big(\mathbf{k}^{(\lambda)}\!\cdot\!\boldsymbol{\sigma}\big)\sigma_x & 0_{2\times 2} & 0_{2\times 2} \\
\gamma\,\sigma_0 & 0_{2\times 2} & M_f(\mathbf \epsilon^{(\eta)}\cdot\boldsymbol\sigma)

\end{pmatrix}.
\label{eq:flat_chiral_ham}
\end{equation}
We find that the single-body Hamiltonian in the unstrained limit ($\epsilon^{(\eta)}=0$) has an additional SU$(2)$ symmetry, attributed to rotations in orbital space. If one considers purely $f-f$ interactions which are rotationally symmetric in the $f$ orbital subspace, such as the on-site Hubbard $U$ considered here, this is an exact symmetry of the many-body Hamiltonian. To the best of our knowledge, this symmetry has not been pointed out before. It has important consequences for the quasi-particle lifetime.

On the single-particle level the symmetry is generated by
\begin{equation}
    \mathcal{O}_i (\mathbf k^{(\eta)}) = \text{diag}(\sigma_i, \mathcal B(\mathbf{k^{(\eta)}})  \sigma_i \big(\mathcal B(\mathbf{k^{(\eta)}}) \big)^\dagger, \sigma_i)
\label{eq:flat_chiral_symmetry}
\end{equation}
with $i=x,y,z$ and 
\begin{equation}
    \mathcal B(\mathbf{k^{(\eta)}}) = \frac{\big(\mathbf{k}^{(\eta)}\!\cdot\!\boldsymbol{\sigma}\big)\sigma_x}{|\mathbf{k}|}.
\end{equation}
The symmetry acts on the many-body wavefunction as 
\begin{equation}
    \Psi_\mathbf k^{(\eta)} \longrightarrow 
    e^{-i\theta_i \hat{\mathcal{Q}}_i^{(\eta)}}
    \Psi^{(\eta)}_\mathbf k e^{i\theta_i \hat{\mathcal{Q}}_i^{(\eta)}}=
    e^{i\theta_i \mathcal{O}_i(\mathbf k^{(\eta)})} \Psi^{(\eta)}_\mathbf k, 
    \label{eq:U1_symmetry}
\end{equation}
With $\theta_i$ the rotation angles around the $x,y,z$ axes, $\Psi^{(\eta)} =(c_1,c_2^,c_3,c_4, f_1, f_2)^T$ the second-quantized annihilation operator in valley $\eta$, and $\hat{\mathcal{Q}_i}^{(\eta)}=\sum_{\mathbf{k}}\Psi^{{(\eta)}^\dagger}_\mathbf{k}\mathcal{O}_i(\mathbf k^{(\eta)} )\Psi^{(\eta)}_\mathbf{k}$ the symmetry charge. Both the kinetic part of the Hamiltonian and the interaction term $U$ in the action \eqref{eq:original_action} are symmetric with respect to \eqref{eq:U1_symmetry}. The interaction part commutes with \eqref{eq:U1_symmetry} because the $f$ part of the symmetry is spatially local. Moving forward we drop the valley indices for brevity.

Notice that a generic rotation $e^{i\theta_i \mathcal{O}_i(\mathbf k)}$ acts non-locally in real-space on the $c$ part. A special sub-symmetry in this regard is the U$(1)$ corresponding to rotations around the z-axis. By noting that $\mathcal{B}(\mathbf{k})$ is unitary and $[\mathcal{B}(\mathbf{k}),\sigma_z]=0$, we see that z-axis rotations do in fact act locally. The associated $\sigma_z$ charge is the relative charge difference between the two chiral sectors, which is already known to be conserved in the chiral limit. 
Crucially, under finite strain the conservation of chiral charge is broken, but a different U$(1)$ subsymmetry is conserved, generated by rotation around the strain axis defined by $\hat{\mathbf{\epsilon}}^{(\eta)}$.

In the strained case, the $f$ mass splits the 8 flat bands into low and high energy sectors whose creation operators we denote by $f^\dagger_-, f^\dagger_+$ respectively. The state created by $f_\pm^\dagger$ acting on the vacuum has eigenvalue $\pm$ with respect to $\mathcal{O}(\mathbf k)\equiv \hat{\mathbf{\epsilon}}_i\mathcal{O}_i(\mathbf k)$. 
We can thus label the two orthogonal sectors of each $\mathbf k$ with $\pm$, defining two projection operators:
\begin{equation}
P_\pm(\mathbf k)\;\equiv\;\frac{1}{2}\Big(\mathbbm{1}\pm \mathcal O(\mathbf k)\Big),\qquad
\Psi_{\mathbf k,\pm}\;\equiv\;P_\pm(\mathbf k)\,\Psi_{\mathbf k}.
\label{eq:projections}
\end{equation}

The associated conserved charge $\mathcal{\hat{Q}}$ together with total particle number conservation results in number conservation in each orthogonal sector, given by:
\begin{equation}
N_\pm\;\equiv\;\sum_{\mathbf k}\Psi_{\mathbf k,\pm}^\dagger \Psi_{\mathbf k,\pm}
\;=\;\sum_{\mathbf k}\Psi_{\mathbf k}^\dagger P_\pm(\mathbf k)\Psi_{\mathbf k}.
\label{eq:sector_numbers}
\end{equation}
We can block diagonalize the single-particle Hamiltonian using this symmetry (see App.~\ref{app:symmmetry}). The resulting sector-resolved non-interacting spectral function, defined as
\begin{equation}
    \label{eq:spectral_func}
    \mathcal{A}(\mathbf k,\omega)=-\frac{1}{\pi}\operatorname{Im}\left[ \operatorname{Tr}G_c(\mathbf k,\omega) + 
    \operatorname{Tr}G_f(\mathbf k,\omega)\right],
\end{equation}
is plotted in Fig.~\ref{fig:ideal}.
As can be seen, at charge neutrality there are two full bands in the $-$ sector and one full band in the $+$ sector. The conservation of particle number $N_\pm$ in each sector has dramatic implications for the problem at all orders in the interaction $U$.

We first consider the limit $\gamma\rightarrow\infty$ where we project to the low energy bands. In this limit the non-interacting ground state is fully polarized Slater-determinant with respect to the U$(1)$ charge $\mathcal{\hat Q}$, having the $-$ sector occupied and the $+$ sector empty. Given that the ground state is fully polarized with respect to $\mathcal{\hat Q}$, because the interaction $U$ respects the symmetry, the exact ground state is the same Slater determinant as the non-interacting ground state. Moreover, any quasi-particle excitation that is charged with respect to $\mathcal{\hat Q}$ is infinitely long lived, because any scattering process must involve a charge conserving particle-hole pair, and there are no such possible pairs. This is reminiscent of QH ferromagnetism, where single quasi-particle excitations of the minority spin are infinitely long lived~\cite{kallin1984excitations}

For finite $\gamma \gg U$ the ground state is no longer a Slater determinant, but low-energy quasi-particle excitations whose $\mathcal{\hat Q}$ charge is different from that of the ground state are still infinitely long lived, the reason being that particle-hole excitations that are neutrally charged with respect to $\mathcal{\hat Q}$ have a gap of order $\gamma$. In other words, because of particle number conservation in each sector there are no on-shell scattering processes for quasi-particle excitations in the active bands. Thus in the flat-chiral limit the flat bands are sharp to infinite order in $U,\,\gamma$. This property of the strained charge neutral case is in sharp contrast to the unstrained case, where in the flat-chiral limit the Mott bands have a width of order $N_f s^2 U$~\cite{vituri2026controlled,hu2026twisted, wei2026lifetime, nosov2026controlled}.

The full THF basis relevant for MATBG has other interaction terms. As explained previously and in App.~\ref{app:interactions}, these interactions will result in a lifetime for the Mott band of order $\sim N_f^2 s^4$. Away from the flat-chiral limit we get corrections at order $N_f s^2$, as discussed below in Sec.~\ref{sec:strained_one_loop}. We can relate these corrections to the deviations that break the symmetry \eqref{eq:U1_symmetry}.

\subsection{Tree-Level Self-Energy}
\label{sec:strained_tree_level}

At charge neutrality, the low energy sector of the $f$-electrons, $f_-$, is full, and the high energy sector $f_+$ is empty. As a result, the complete, non-perturbative effect of the interaction for a single-particle excitation is given by the Hartree-Fock correction $M_f \rightarrow M_f + u$ for the single-particle Hamiltonian, where we defined $u\equiv\tfrac{U}{2}$.   
For the hybridized problem, this means the Hubbard-I propagator has the form $\left(i\omega-\tilde h_\mathbf k ^{(\eta)}\right)^{-1}$, where $\tilde h_\mathbf k ^{(\eta)}$ is given by \eqref{eq:thf_hamiltonian} with the substitution $M_f(\mathbf \epsilon^{(\eta)}\cdot\boldsymbol\sigma) \rightarrow \left(M_f|\mathbf \epsilon^{(\eta)}| +u\right)(\mathbf{\hat \epsilon^{(\eta)}}\cdot\boldsymbol\sigma)$, with $\mathbf{\hat \epsilon^{(\eta)}}$ the direction of the vector $\mathbf \epsilon^{(\eta)}$.

\begin{figure*}
    \centering
    \includegraphics[width=\linewidth]{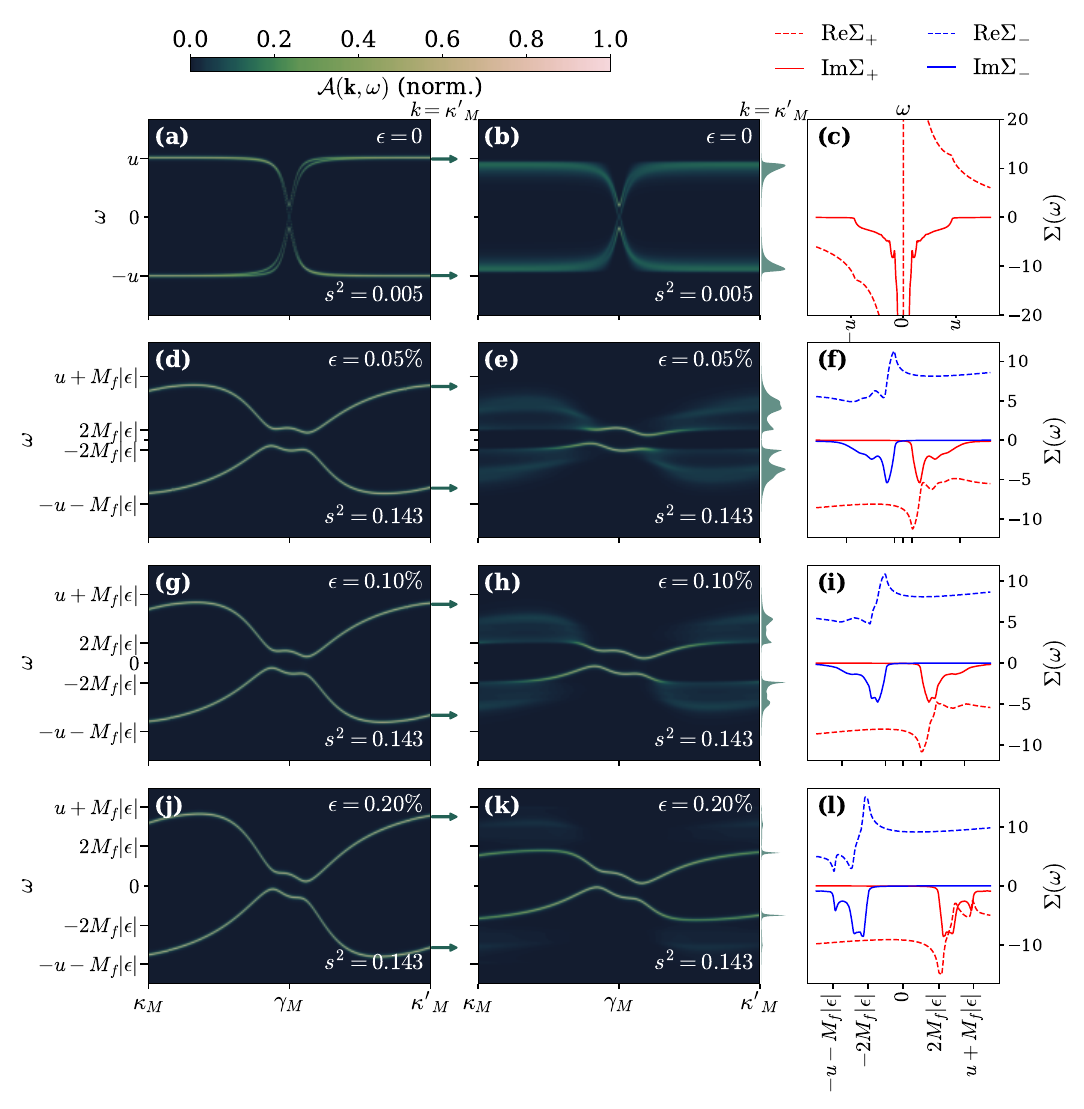}
    \caption{(a) Spectral function $\mathcal{A}(\mathbf k,\omega)$ along the $\kappa_M$-$\gamma_M$-$\kappa'_M$ line calculated to tree-level for the unstrained case. (b) Unstrained spectral function to one-loop order, (c) the self-energy to one-loop order. To the side of panels (a), (b) we show the spectral cut $\mathcal{A}(\kappa'_M,\omega)$, where for (a) the delta functions of the tree-level spectrum are indicated with arrows. Both calculations are done as explained in Sec.~\ref{sec:controlled_expansion_C} using Eq.~\eqref{eq:S_definition}, \eqref{eq:f_self_energy}, \eqref{eq:c_self_energy}. Panels (d)-(f), (g)-(i), (j)-(l) show the same results for the cases of $0.05\%$, $0.1\%$, $0.2\%$ strain, respectively, with uniaxial heterostrain angle $\phi=15^\circ$. For the unstrained case there is no high or low energy sector and thus $\Sigma=\Sigma_+=\Sigma_-$. All parameters are the same for all graphs except for the strain magnitude and the use of a larger value of $v_\star$ for the unstrained case, as explained in Table \ref{table:params}. The use of a larger $v_\star$ results in $s^2$ being smaller for the unstrained case. The strained spectrum with a larger $v_\star$ is presented in Fig.~\ref{fig:app_spec} in App.~\ref{app:extra}.}
    \label{fig:spectrum}
\end{figure*}

There are a few spectral features unique to the strained case that are worth mentioning.
First, it is worth noting that because this is effectively a single electron problem, we have 6 bands in the spectrum per valley, per spin, each with spectral weight 1. The strain breaks the rotational invariance of the spectrum, with the Mott band energy now dependent on the $k$ angle. Moreover, the spectrum has two Dirac nodes, whose locations and energies are not fixed to the $\gamma_M$ point, but depend on the strain angle and magnitude. A detailed calculation showing these features is presented in App.~\ref{app:tree_level}, and the tree-level spectrum is presented in Fig.~\ref{fig:spectrum}, with the spectral function defined as \eqref{eq:spectral_func}. The spectral function of strained MATBG was previously calculated in the Hubbard-I approximation in Ref.~\cite{ledwith2024nonlocal}, and at finite temperature in Ref.~\cite{ledwith2025exotic}. It was also calculated using Hartree-Fock~\cite{Kang2020,liu2021nematic, xie2020nature, bultinck2020ground, parker2021strain, kwan2021kekule, herzog2025kekule}, which gives the correct tree-level correction as discussed here.

\subsection{One-Loop Order Self-Energy}
\label{sec:strained_one_loop}
To calculate the $f$ self-energy to one-loop order we follow the method outlined in Sec.~\ref{sec:controlled_expansion_C}. We start by deriving the long-range in imaginary time part of $\Gamma^{(2)}$, and use it to calculate Eq.~\eqref{eq:S_definition} to one-loop order. Using Eq.~\eqref{eq:f_self_energy},\eqref{eq:c_self_energy} and \eqref{eq:D_S_relation} we can then obtain the $f$ self-energy, and the corresponding $c$ self-energy. Note that to order $s^2$, the one-loop correction in Eq.~\eqref{eq:f_self_energy} is simply given by $S$. Finally, we plot the resulting spectral function in Fig.~\ref{fig:spectrum} and discuss key features in it, including the appearance of a trion resonance unique to the strained case. Additional figures, highlighting the parametric dependence of the spectral features in the strained case and contrasting between the strained and unstrained cases are presented in App.~\ref{app:extra}.

\subsubsection{$f$ 4-point function}
We consider the 4-point function,
\begin{equation}
    \big\langle 
f_{\lambda_1,b_1}(\tau_1) f_{\lambda_2,b_2}(\tau_2)
\bar f_{\lambda_2,b_2}(\tau_2') \bar f_{\lambda_1,b_1}(\tau_1')
\big\rangle_{c,0},
\label{eq:4_point_func}
\end{equation}
in the regime $u \gg M_f|\mathbf \epsilon| \gg T$, where we can define a long-time scale
\begin{equation}
\tau_{\text{long}} \sim \frac{1}{M_f|\mathbf \epsilon|} \gg \frac{1}{u}.
\label{eq:parameters_limits}
\end{equation}
Following Ref.~\cite{vituri2026controlled} we focus on long-range in time processes. At charge neutrality, there is no possibility for the zero-energy flavor-flips found in the unstrained case. However, we can focus on ``orbital-flip" processes where we create a hole in the low energy sector $f_-$ and an electron in the high energy sector $f_+^\dagger$. In the strained single-site problem, this orbital-flip excitation has energy $2 M_f |\mathbf \epsilon|$, and will propagate for $\sim\tau_\text{long}$ before annihilating.  
For \eqref{eq:4_point_func} to be non-zero one pair of operators must be in the $+$ sector and the other in the $-$ sector. For example, taking $b_1=+, b_2=-$, one time ordering gives:
\begin{widetext}
\begin{equation}
\begin{split}
        \big\langle f_{\lambda_1,+}(\tau_1)\,\bar f_{\lambda_2,-}(\tau_2')&\,f_{\lambda_2,-}(\tau_2)\,\bar f_{\lambda_1,+}(\tau_1')\big\rangle_{c,0} = \\
=
&-\,\Theta(\tau_2-\tau_1')\,
  \Theta(\tau_2'-\tau_2)\,
  \Theta(\tau_1-\tau_2')\,
\times
\exp\!\Big[-(u+M_f|\mathbf \epsilon|)(\tau_2-\tau_1'+\tau_1-\tau_2')
           -2M_f|\mathbf \epsilon|(\tau_2'-\tau_2)\Big]
\end{split}
\end{equation}
The contribution from $b_1=-, b_2=+$ is exactly the same after renaming the time variables. The total result for the long-range in time part of \eqref{eq:4_point_func} in Matsubara frequencies is then:
\begin{equation}
\begin{split}    
\big\langle
f_{b_1,i\omega_1}& f_{b_2,i\omega_2}
\bar f_{b2,i\omega_2'} \bar f_{b1,i\omega_1'}
\big\rangle_{c,0}
=
\beta\,\delta_{\omega_1+\omega_2,\;\omega_1'+\omega_2'}\,
(1-\delta_{b_1,b_2}) \times\\
& \times\Big[\delta_{b_1,+}\Big(
\frac{1}{(u+M_f)-i\omega_1'}+\frac{1}{(u+M_f)+i\omega_2}
\Big)
\Big(
\frac{1}{(u+M_f)-i\omega_1}+\frac{1}{(u+M_f)+i\omega_2'}
\Big)
\frac{1}{2 M_f |\mathbf \epsilon | - i(\omega_1-\omega_2')}
\\
&\qquad\qquad\qquad\qquad\qquad\qquad\qquad\qquad\qquad\qquad\qquad\qquad\qquad\qquad
+\delta_{b_1,-}(\omega_1\leftrightarrow\omega_2, \omega_1'\leftrightarrow\omega_2')
\Big]
\label{eq:4_point_func_result}
\end{split}
\end{equation}
\end{widetext}

We ignore other contributions to $\Gamma^{(2)}$ such as the ones coming from charge fluctuations, as these are short-range in time and we expect them to be irrelevant for the lifetime of the Mott band.

\subsubsection{One-loop diagram}

To first order, Eq.~\eqref{eq:S_definition} is given by
\begin{align}
&\hat\gamma^\dagger G_{f,0} S^{(1)}_\lambda(i\omega) G_{f,0} \hat\gamma
=
\nonumber \\
&\sum_{\lambda'}
\int_{\mathbf{k}'}\int_{\omega'} G^{0}_{c,\lambda'}(\mathbf{k}',i\omega')\,
\Gamma^{(2)}_{\lambda, \lambda'}(\omega,\omega', \omega', \omega),
\label{eq:one_loop_diag_expression}
\end{align}
where $\int_{\mathbf{k}'}\equiv\int_{BZ} \frac{d^2k'}{A_{BZ}}$, $\int_{\omega'}\equiv\int_{-\infty}^{\infty}\frac{d\omega'}{2\pi}$. As shown in Appendix \ref{app:one_loop}, $S^{(1)}$ is diagonal in the basis that diagonalizes the strain term $\mathbf{\epsilon}\cdot \mathbf{\sigma}$. Writing $S^{(1)}$ in that basis, we get:
\begin{equation}
\begin{aligned}
    S^{(1)}(i\omega) = 
    \begin{pmatrix}
        S^{(1)}_+(i \omega) & 0 \\
        0 & S^{(1)}_-(i \omega)
    \end{pmatrix},
\end{aligned}
\end{equation}
with 
\begin{align}
    &S^{(1)}_\pm(i\omega)= \nonumber \int_{\omega'}
\mathcal{ I}_\mp(i\omega') \times\\
    &\qquad \times\left[1+\frac{(u+M_f|\mathbf \epsilon|)\mp i\omega}{(u+M_f|\mathbf \epsilon|) \pm i\omega'}\right]^2 \frac{1}{2M_f|\mathbf \epsilon| \pm i(\omega'-\omega)}.
    \label{eq:S_pm}
\end{align}
In the above expressions, $\mathcal I_\pm$ denotes the diagonal elements of the matrix in the strained basis, with $\mathcal I$ defined in \eqref{eq:I_def}. The subscript $\pm$  corresponds to the high and low energy orbital, respectively.
As shown in App.~\ref{app:one_loop}, we can get a more explicit result for the imaginary part in the flat bands:
\begin{align}
    \operatorname{Im} S_\pm^{(1)}(\omega) &= -\pi\theta(\pm\omega - 2M_f|\mathbf \epsilon|)\nonumber\\ 
    &\times
    \rho_\mp(\omega\mp2M_f|\mathbf \epsilon|)
    \left[\frac{2u}{u-M_f|\mathbf \epsilon|\pm\omega}
    \right]^2,
\label{eq:S_imag}
\end{align}
with $\theta(\omega)$ the Heaviside function and
\begin{equation}
    \rho_\pm(\omega)
=
-\frac{1}{\pi}\,\operatorname{Im} \mathcal{I_\pm}(\omega + i0^+)
\end{equation}
the density of states of the $c$ particles in the loop attached to the $\pm$ orbital, up to conjugation with $\hat \gamma \hat \gamma^\dagger$. The condition  $\rho_\pm(\omega)\ll \omega$ justifies the perturbative loop expansion. As seen in App.~\ref{app:one_loop}, in the flat-chiral limit $\rho_\pm \propto N_f s^2 \omega$ with $s^2$ the parameter defined in \eqref{eq:s2_def}. In addition, in the flat-chiral limit $ \rho_\pm$ vanishes for negative and positive frequencies, respectively, resulting in the imaginary part \eqref{eq:S_imag} vanishing, as required by the symmetry presented in \ref{sec:strained_symmetry}.

The Heaviside function is a standard feature of the self-energy for inelastic scattering off a constant energy bosonic mode, as explained further in the next section, and pointed out in Ref.~\cite{hu2026twisted}. To one-loop order the $f$ self-energy is then:

\begin{equation}
    \Sigma_f = \hat \gamma G_c \hat \gamma^\dagger+S^{(1)}.
\end{equation}

\subsubsection{Spectral features at one-loop order}
\label{sec:strained_features}
In Fig.~\ref{fig:spectrum} we plot the spectrum of strained MATBG for a range of strain magnitudes, with additional plots for different model parameters found in Fig.~\ref{fig:app_grid}. We now analyze the main differences compared to the tree-level spectrum. The new spectral features can be understood from Eq.~\eqref{eq:S_imag}

First, the factor of $ \rho_{\pm}$ results in a vanishing imaginary part at the flat-chiral limit. We can qualitatively explain the effect of each of the symmetry breaking parameters on the Mott bands' width, knowing that  $ \rho_{\pm}(\mp u \pm M_f |\mathbf \epsilon|)$ controls it.

The effect of $M$ can be neglected away from the $\gamma_M$ point where $|\mathbf{k}|\gg\frac{M}{v_\star}$, and so we expect it to be irrelevant to $ \rho_{\pm}(\mp u \pm M_f |\mathbf \epsilon|)$ and therefore to the lifetime. We can expect $c''$ to add a contribution to the lifetime of order $s^2 (\frac{c''|\mathbf \epsilon|}{\gamma})^2$, which is $\sim 2$ orders of magnitude smaller than $s^2$ when using standard values of $\gamma, c''$ and $|\mathbf \epsilon| \lesssim 0.1\%$~\cite{cualuguaru2023twisted}.  

For standard model parameters, we therefore expect $v_\star'$ to be the most important parameter for the lifetime of the Mott band. As $v_\star'$ is also closely related to the ratio $\frac{w_0}{w_1}$~\cite{cualuguaru2023twisted} this gives an important connection between the distance from the chiral limit and the width of the Mott bands in the strained case. 

Lastly, we expect $ \rho(\omega)$ to increase with $|\omega|$, since the tree-level density of states (DOS) increases linearly away from the Mott band. This means $ \rho_{\pm}(\mp u \pm M_f |\mathbf \epsilon|)$ and therefore \eqref{eq:S_imag} decrease with increasing strain. All these qualitative arguments are in agreement with the numerical calculations obtained for the spectral function and self-energy, as can be seen in Figs.~\ref{fig:spectrum},~\ref{fig:app_grid}. The large difference between the Mott bands' width in the strained compared to the unstrained case is highlighted in Fig.~\ref{fig:app_spec}, where for the same model parameters the flat bands' width is negligible in the strained case when compared to the unstrained case.

The most distinct spectral feature of the strained spectrum is the new flat band that emerges at or below $2 M_f |\mathbf \epsilon|$. This spectral feature appears because of the inelastic scattering from a constant energy bosonic mode, in this case the orbital-flip particle-hole pair created by $f_+^\dagger f_-$, but it is easiest to understand by analogy to the Holstein model of electron-phonon coupling~\cite{marsiglio2008electron}, which we go through in App.~\ref{app:hosltein}. In the electron-phonon problem, the electron's self-energy at one-loop order is given by:

\begin{equation}
\Sigma(\omega + i0^+) =
\lambda \omega_E
\left[
\ln \left| \frac{\omega_E - \omega}{\omega_E + \omega} \right|
- i \pi \, \theta(|\omega| - \omega_E)
\right] ,
\label{eq:holstein_self_energy}
\end{equation}
with $\lambda$ the dimensionless electron-phonon coupling and $\omega_E$ the Einstein phonon frequency. As a result of Kramers-Kronig relations, such a sharp feature in the imaginary part of the self-energy always results in the divergence of the real part. The poles of the electron's Green's function are then given by:
\begin{equation}
\omega = \epsilon_k + \rm{Re} \Sigma(\omega) .
\label{eq:holstein_poles}
\end{equation}
and because of the logarithmic divergence of $\rm{Re} \Sigma(\omega)$ at $\omega_E$ there is a pole below $\omega_E$ at any $k$ for which $\epsilon_k>\omega_E$. As shown in App.~\ref{app:hosltein}, the residue on this pole is proportional to $\exp\!\left(\frac{-\epsilon_k/\omega_E}{\lambda}\right)$ with $\epsilon_k$ the band dispersion, thus for a generic $\epsilon_k$ this gives the known ``kink" in the spectrum and there is no visible pole below $\omega_E$, as the residue vanishes exponentially. In the case of a flat dispersion, the spectral weight on this extra pole can be substantial, as $\epsilon_k/\omega_E$ is of order 1 for all $k$.

Going back to the case of strained MATBG, we expect a similar feature as a result of the Heaviside function in the imaginary part of the self-energy \eqref{eq:S_imag}. In this case, the flat dispersion allows for a band to appear as a result of the new solution of Eq.~\eqref{eq:holstein_poles} below $2 M_f |\mathbf \epsilon|$. 

The appearance of this feature can also be understood differently. In the non-hybridized limit, there are excitations of the form $c^\dagger_{\lambda_1, a} f^\dagger_{ \lambda_2, +} f_{\lambda_3, -}$ with an energy 
\begin{equation}
 2 M_f |\mathbf \epsilon| + \varepsilon_{c,\mathbf k},   
 \label{eq:trion_energy_like}
\end{equation}
with $\varepsilon_{c,\mathbf k}$ the $c$-electron single-particle energy. When we add hybridization, these states mix with the single-particle excitations as a result of the $U$ interaction term, and we expect them to appear as a new feature in the single-particle spectrum around the energy $2M_f |\mathbf \epsilon|$. We thus relate these strain-induced trion excitations with the new band appearing around $2M_f |\mathbf \epsilon|$, and we call this new band the trion band.\footnote{Note that the trion we discuss here is different from the one identified in Ref.~\cite{ledwith2025exotic}, as our trion is related to the orbital-flip in the strained case, and appears in the spectral function away from the $\gamma_M$ point, unlike the one in Ref.~\cite{ledwith2025exotic}. We further note that the trion band discussed here is unrelated to the extra feature seen in Fig.~2(c) of Ref.~\cite{ledwith2025exotic}, which is a result of the finite temperature filling of the lower and higher energy $f$-orbitals.}

The trion band is located at or below the energy $2 M_f |\mathbf \epsilon|$. Because the one-loop lifetime \eqref{eq:S_imag} is infinite below the onset energy $2 M_f |\mathbf \epsilon|$, the trion band is sharper than the Mott band and has a width of order $s^4$. 

We can expect the energy of the strain-induced trion to decrease when the available phase space for excitations of the form $c^\dagger f^\dagger f$ increases. The relevant excitations must have the same quantum numbers as a single-particle excitation, and as seen in Sec.~\ref{sec:strained_symmetry} there is no such available phase space in the flat-chiral limit because of the lack of particle-hole excitations that are neutral with respect to the charge $\mathcal{\hat Q}$. For example, $f_+^\dagger$ can mix with excitations of the form $c_-^\dagger f_+^\dagger f_-$, and at the flat-chiral limit $c_-^\dagger$ annihilates the ground state (if we ignore the remote bands). This fact, along with the expectation that lower energy particles are more important to the strain-induced trion because of the factor $\varepsilon_{c,\mathbf k}$ in \eqref{eq:trion_energy_like}, suggests that breaking the symmetry \eqref{eq:U1_symmetry} close to $\omega=0$ lowers the trion's energy significantly. In other words, the important quantity is the $c$-electrons DOS in the opposite sector close to $\omega=0$. For small, positive energies, this is the DOS of the $c$-electrons in the $-$ sector. In terms of the model parameters, this means that increasing $c''|\mathbf \epsilon|$, which shifts the energy of the Dirac node, and to a lesser degree $M$, will increase the available phase space for trion-single-particle mixing, and thus lower the energy of the new trion band. We further note that the remote bands host excitations with the opposite charge with respect to $\mathcal Q$ (see Fig.~\ref{fig:ideal}), and thus mixing with the remote bands will lower the energy of the new trion band. These expectations are in agreement with the numerical results.

We further note that in the electron-phonon toy model the ratio $\frac{\epsilon_k}{\omega_E}$ controls the spectral weight of the trion feature. In strained MATBG, this corresponds to $\sim \frac{u+M_f|\mathbf \epsilon|}{2M_f|\mathbf \epsilon|}$. Since this quantity decreases for larger $|\mathbf \epsilon|$ we expect more spectral weight on the trion band at larger strain values. This is also in agreement with the numerical results, as seen in Fig.~\ref{fig:spectrum} and Fig.~\ref{fig:app_grid}. 

Lastly, we note that the location and the relative spectral weight on the trion feature depends on the real part of the self-energy, including the non-divergent parts. These include contributions from all the remote bands, which depend on the cutoff, as well as the short-range in time part of the vertex $\Gamma^{(2)}$. Taking into account these contributions can change the quantitative nature of the trion for a specific set of model parameters. Nevertheless, we stress that the strain-induced trion is always a feature of the model, and by varying the other model parameters it can become more or less pronounced. As can be seen in Fig.~\ref{fig:app_grid}, for some model parameters most of the spectral weight shifts from the Hubbard band to the trion band away from the $\gamma_M$ point.

\section{QTM Spectrum}
\label{sec:qtm}
The spectral function of MATBG was recently measured in a momentum- and energy-resolved measurement using a Quantum Twisting Microscope (QTM) for the first time~\cite{inbar2023quantum, xiao2025interacting}. In this section we briefly go through the theoretical background of the QTM measurement~\cite{wei2025dirac}. Afterwards, we present and compare the theoretically predicted QTM spectrum in the strained and unstrained cases at charge-neutrality in the hope of distinguishing between the two experimentally.

\subsection{Theoretical Background}
\label{sec:qtm_theory}
We briefly recap the theory behind the QTM measurement following Ref.~\cite{wei2025dirac}. For a complete review we point the reader to that reference.

In the QTM setup, a van der Waals tip, usually monolayer graphene (MLG), is used to probe local properties of another van der Waals sample, such as MATBG, by forming a twistable finite-area tunneling junction. The finite area of the junction results in a coherent tunneling process that is momentum and energy resolved. This enables the QTM to measure the momentum-resolved energy dispersion of the sample device.

The equation for the current through the QTM device is given by:

\begin{widetext}
\begin{equation}    
I = \frac{2\pi e}{\hbar} 
\int d\omega \,\big[ f(\omega) - f(\omega + eV_b) \big]
\sum_{k,k'} \sum_{\alpha,\alpha'}
\big|\langle k'\alpha'_{T} \,|\, H_{\rm tun} \,|\, k\alpha_{S} \rangle\big|^2 
A^T_{\alpha'}(k',\omega+eV_b)\,A^S_{\alpha}(k,\omega),
\end{equation}
\end{widetext}
where $f(\omega)$ is the Fermi distribution, $A^T$ and $A^S$ are the spectral functions of the tip and sample, respectively, and $\alpha,\alpha'$ label different bands, $H_{tun}$ is the tunneling Hamiltonian between the tip and the sample and $V_b$ is the bias voltage. The sharpness of the Fermi surface and the singularity of the density of states of the MLG tip at the Dirac point result in sharp features in the derivatives $\frac {d^nI} {(dV_b)^n}$ when $V_b$ crosses a band of the sample. These features appear when 
\begin{equation}
eV_b + \mu_T + \xi^{S,\alpha}_{p} = 0,
\end{equation}
with $\mu_T$ the tip's chemical potential, $\xi^{S,\alpha}_{p}$ the energy dispersion of the sample for the band $\alpha$ at momentum $p$, relative to the sample's chemical potential. 

\begin{figure*}
    \centering
    \includegraphics[width=\linewidth]{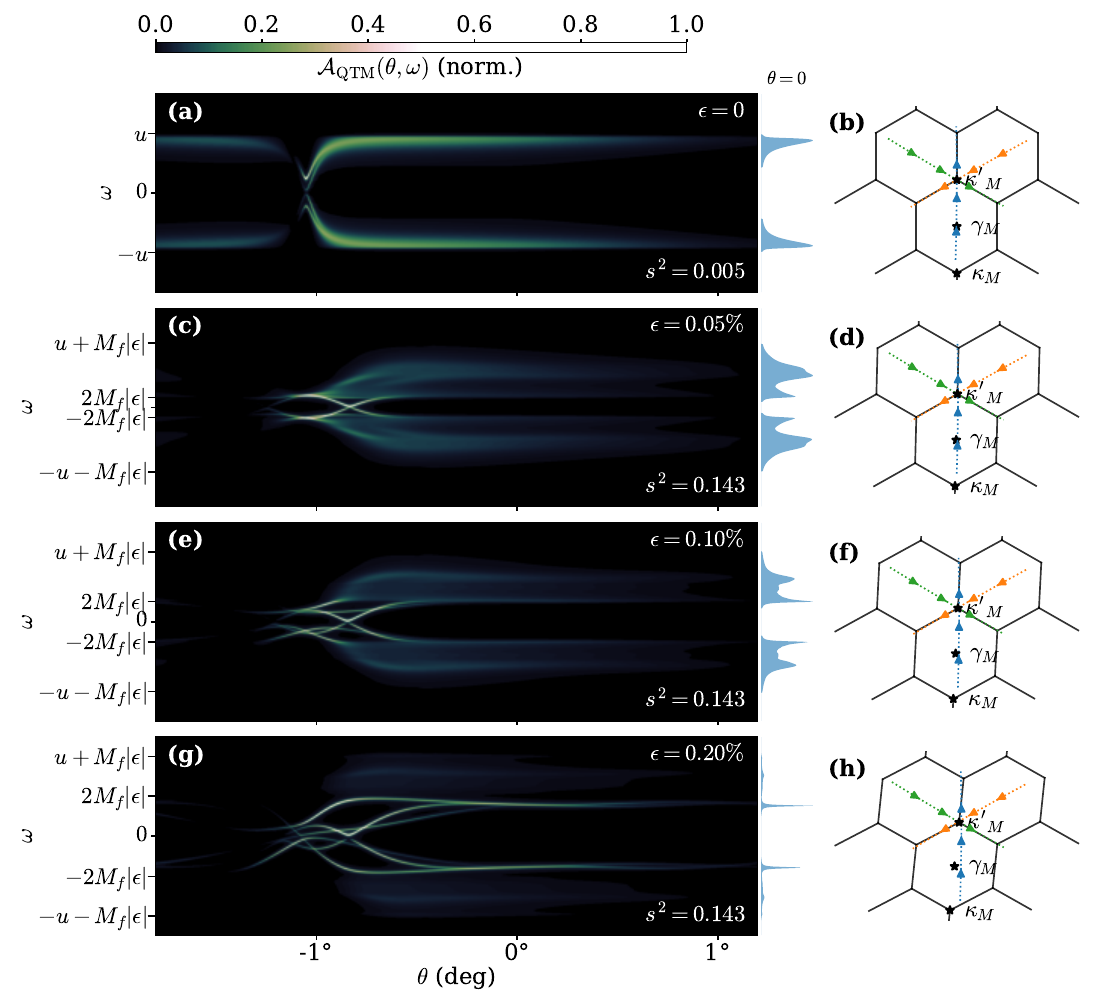}
    \caption{(a) QTM spectrum calculated using the spectral function to one-loop order for the unstrained case. The result is the sum of the six paths scanned by the three $K_n$ points in the two valleys. To the right of the panel we plot the QTM spectral weight cut at $\theta=0^\circ$, normalized relative to its maximal value. (b) Paths scanned by the QTM in the $+$ valley, each arrow shows a path corresponding to one of the $K_n$ points of the tip. The paths in the other valley are related to those shown here by time reversal symmetry. Here $\theta=0^{\circ}$ corresponds to alignment of the QTM tip with the point $\kappa'_M$. (c)-(d), (e)-(f), (g)-(h) show the same results for the cases of $0.05\%$, $0.1\%$, $0.2\%$ strain, respectively, with uniaxial heterostrain angle $\phi=15^\circ$. The parameters are the same as in Fig.~\ref{fig:spectrum}, with all parameters being the same for all graphs expect the strain magnitude and a different $v_\star$ value in the unstrained graph.}
    \label{fig:qtm}
\end{figure*}

Because the tunneling occurs through a planar junction momentum is conserved up to reciprocal lattice vectors of the tip and scanned sample. The rotation of the QTM results in different sample momenta satisfying the momentum conservation for different twist angles, which allows scanning the spectral function along a line. 
For a MLG tip with $\mu_T=0$ the sample's spectral function is scanned along six different lines simultaneously as a result of the two-fold valley and three-fold $K$ point degeneracy of the MLG Dirac cones. Time reversal symmetry results in the spectrum being the same along the scanned lines for both valleys, and we label the different $K$ lines with $n=1,2,3$. These lines are sketched in Fig.~\ref{fig:qtm}(b),(d),(f),(h) for different strain values. For the unstrained case the spectrum is the same along the different $n$ lines as a result of $C_{3z}$ symmetry, while in the strained case the spectrum is different along each line, as already seen in calculations for MATBG using the BM model~\cite{wei2025dirac}.

The location and magnitude of these features were theoretically calculated for $\frac {d^2I} {(dV_b)^2}$ in Ref.~\cite{wei2025dirac}, while experimentally $\frac {dI} {dV_b}$ was measured in Ref.~\cite{xiao2025interacting}. In this work we plot
\begin{equation}
    \frac{dI}{dV_b}\propto 
    \sum_{\alpha, \eta, n} A^S_\alpha(\mathbf k, \omega) \left|e^{i\eta \frac{2\pi(n-1)}{3}}\psi^\alpha_{tA} + \psi^\alpha_{tB} \right|_{\mathbf K_{\eta,n}},
\label{eq:QTM_spectrum}
\end{equation}
which captures the right singularities structure and intensity of the spectrum for both cases. In addition to the spectral function, the second term is a result of the tunneling matrix element from the tip to the sample, with $\psi^\alpha_{tA/B}$ the wavefunction of an excitation in the band $\alpha$ on the $A/B$ site of the top layer, and $\mathbf K_{\eta,n}$ denotes the fact that this is evaluated at the momentum scanned along the line $n$ in the valley $\eta$~\cite{wei2025dirac}. This expression qualitatively reproduces the QTM spectrum calculated for the BM model in Ref.~\cite{wei2025dirac} and seen experimentally in Ref.~\cite{xiao2025interacting} away from the magic angle, as shown in App.~\ref{app:bm_qtm_spectrum}.

\subsection{Unstrained vs. Strained MATBG}
\label{sec:qtm_comparing}

The QTM spectrum is calculated using Eq.~\eqref{eq:QTM_spectrum} assuming a MLG tip with $\mu_T=0$, with the results presented in Fig.~\ref{fig:qtm}. The unstrained spectrum is calculated as explained in Ref.~\cite{vituri2026controlled}, with the one-loop self-energy calculated as explained in Sec.~\ref{sec:controlled_expansion_C} of this paper. For more details about the numerical calculation, see App.~\ref{app:qtm_numerics}

Comparing the strained and unstrained cases, the most significant difference is the number of scanned lines. As explained in Ref.~\cite{wei2025dirac} and Sec.~\ref{sec:qtm_theory}, the strained case breaks $C_{3z}$ and as a result the three scanned $K_n$ lines scan different spectra. We thus expect three different spectral lines to appear in the QTM spectrum of the strained case.

In addition, differences in the spectrum of the strained and unstrained cases that appear already at tree-level can be seen in the QTM spectrum. The band touching point, which is fixed to the $\gamma_M$ point in the unstrained case, is replaced with Dirac cones away from the $\gamma_M$ point in the strained case. The locations of the Dirac nodes depend on the angle and magnitude of the strain, but the resulting spectrum as seen by the QTM is nevertheless different compared to the unstrained case, as seen in Fig \ref{fig:qtm}. 

Other features that differ between the strained and unstrained cases include the appearance of the kink or trion band around the energy $2 M_f |\mathbf \epsilon|$. In some parametric regimes, we therefore see an extra band in the strained case, with one band's energy around $u+M_f|\mathbf \epsilon|$ and the other around $2M_f|\mathbf \epsilon|$, compared with the unstrained case where there is a single flat band with an energy $u$. Furthermore, for the same set of parameters, we find a very different width of the Mott bands for the strained and unstrained cases. In the unstrained case the width is of order $N_f s^2 u$~\cite{vituri2026controlled,hu2026twisted,wei2026lifetime, nosov2026controlled}, while in the strained case it is of order $s^4$ in the flat-chiral limit, with corrections being parameter dependent but substantially smaller than in the unstrained case. The strain-induced trion band is sharp to order $s^2$ away from the flat-chiral limit as well.

We note that both the new spectral features unique to the strained case, as well as the width of the Mott bands are parameter dependent features. It is therefore easier to use them as constraints on theoretical parameters when comparing experiment to theory than it is to use them as definite proof to conclude whether a sample is strained or not. We further note that the finite area of the QTM tip results in an uncertainty in $k$ on the order of $\sim\frac{1}{L}$ in measurements, with $L$ the size of the tip. This uncertainty can blur certain strained MATBG spectral features.

\section{Discussion \& Outlook}
\label{sec:discussion}

In this work we studied MATBG under strain in the appropriate strong-coupling regime using a small-hybridization-phase-space loop expansion~\cite{vituri2026controlled}. In order to do so, we developed a systematic way to perform a proper Dyson resummation for the low-energy $f$ degrees of freedom, which are traced out within the current formalism. This is done by identifying strong-coupling diagrams in the effective field theory with infinite sets of diagrams in the full theory, formalized as a perturbative expansion in~$U$. 

Our results are complementary to DMFT studies of strained MATBG~\cite{crippa2025dynamical}. The main advantage of the current approach, in addition to the analytical control, is that the calculations can be done directly in real frequency, without the need for analytical continuation of noisy numerical data. As a result, we are able to resolve fine spectral features, such as the strain-induced trion band (see below).

We found that at charge-neutrality,  under significant strain, the scattering of low-energy electrons is dominated by an inelastic process described by an $f$ site orbital-flip. The energy loss in this process equals $2M_f|\mathbf \epsilon|$ -- the strain-induced orbital splitting energy of the $f$ site. This is in contrast to the unstrained case, where the analogous scattering is elastic. 

We further identified an additional approximate symmetry of the model, having significant consequences in the strained case. In the flat-chiral limit, with only orbitally symmetric $f$-$f$ interactions, this symmetry is exact. Then, if the band gap is sufficiently large, strained MATBG at $\nu=0$ has infinite quasiparticle lifetime at any loop order, by analogy to quantum Hall ferromagnetism. As a result we find that over a wide range of realistic model parameters, where the symmetry is only approximate, the addition of strain reduces the Mott bands' width by more than an order of magnitude. 

In addition, we found a new excitation, unique to the strained case. This excitation, having electron or hole-like quantum numbers, is a strain-induced trion. It is composed of a bound state between an $f$ site exciton and a low-energy electron or hole. The $f$ site exciton in this case is a particle-hole pair of an hole from the occupied (low-energy) strain sector and a particle from the unoccupied (high-energy) sector. The exciton has an energy of $2M_f|\mathbf{\epsilon}|$, whereas the bound trion has a strictly lower energy. The binding energy of the strain-induced trion increases as we increase the amount by which the aforementioned symmetry is violated at low energies. As the trion energy is given, to leading order in $s^2$, by the neutral exciton energy, we do not expect its energy to depend strongly on chemical potential. That is in contrast to a single-particle excitation whose energy depends linearly on the chemical potential.

Another direct consequence of the inelastic scattering is that to one-loop order (i.e. to order $s^2$), quasi-particles at energies below $2M_f|\mathbf \epsilon|$ do not scatter. This statement is robust over a finite range of densities around $\nu=0$ -- namely, the range of densities between the first compressibility peaks (commonly referred to as cascades) at positive and negative densities. This leads us to the following experimental prediction: Upon doping the system around $\nu=0$ certain spectral features, including the Mott bands, shift in energy relative to the Fermi level ($\omega_\text{Mott}=\pm \frac{U}{2}-\mu$). As these features cross below the energy threshold for scattering $|\omega|<E_\text{inelastic}$, scattering processes are highly suppressed and appear only at higher order. Therefore, we predict a significant sharpening of all spectral features at a certain well-defined energy we dub $E_\text{inelastic}$. This prediction applies to any spectral feature, but is expected to be visible most clearly in the Mott bands away from the $\gamma_M$ point, where the band is flat and this effect is the leading effect upon doping. 
Such observation will not only verify the existence of a significant strain, but also give a direct measurement of the low-energy strain-induced energy scale $E_\text{inelastic}$. 
We note that $E_\text{inelastic}=2M_f|\mathbf  \epsilon| + O(U s^2)$; in reality, $E_\text{inelastic}$ may be significantly renormalized relative to $2M_f|\mathbf  \epsilon|$. In addition, by the analogy to the electron-phonon problem, as the Mott band moves below $E_\text{inelastic}$ we expect the trion band to vanish (see discussion below Eq.~\eqref{eq:holstein_poles}).

Lastly, we showed the predicted QTM spectrum in the strain and unstrained cases. As seen in Sec.~\ref{sec:qtm}, all the spectral features unique to the strained case are expected to be visible in the QTM spectrum. 
In reality the momentum resolution of the QTM is limited by the inverse linear tip size. 
For a tip size of $\sim100$nm, the finite momentum resolution can blur the highly detailed dispersive features near the $\gamma_M$ point. Flat features such as the spectral kink, the trion band appearing below $2M_f|\mathbf \epsilon|$, and the Mott band are less sensitive to momentum resolution, and should be easily distinguishable in QTM and in STM experiments. 
Note that the fact that only two flat bands were seen in Ref.~\cite{xiao2025interacting} does not preclude the existence of significant strain. As seen clearly in Fig.~\ref{fig:qtm}, there are parametric regimes where the trion bands are the only flat bands visible in the spectrum. The other extreme, where the Hubbard bands are the only visible flat bands, can also be seen in Fig.~\ref{fig:app_grid}, in which case we expect to see a kink in the spectrum at $2M_f|\mathbf \epsilon|$. Depending on the strain magnitude and the value of $s^2$, both regimes can explain the existence of only two flat bands. 

This work, demonstrating significant qualitative difference between strained and unstrained MATBG at $\nu=0$ leaves a few open questions which we intend to answer in future works. A first and immediate one is how this difference translates to non-zero fillings (both integer and non-integer ones). We emphasize the uniqueness of $\nu=0$ in the strained case, having a unique Slater determinant ground state of the single-site problem. At finite fillings this is not the case, as the single-site ground state is degenerate and allows for elastic scattering processes akin to those found for unstrained MATBG~\cite{vituri2026controlled, hu2026twisted, wei2026lifetime, hu2026twisted}, in addition to the inelastic ones found at charge neutrality. Away from charge neutrality, we thus expect the width of one of the Mott bands to increase. For example, for integer $\nu>0$ the electron-like Mott band will have a larger width compared to charge neutrality, for the same reasons that the unstrained case has a larger width compared to the strained case at charge neutrality. For the same reason, at finite fillings, similar to the unstrained case, the expansion is valid only at intermediate temperatures below which the system is expected to break symmetry spontaneously or to arrive at a strongly fluctuating critical regime.

A second extension is to consider an intermediate scenario between spontaneous symmetry breaking and an explicit symmetry breaking governed by an external energy scale. In this work we explicitly focused on the case of large strain such that $M_f|\mathbf{\epsilon}|\gg\Theta_c$ ($\Theta_c$ being the Curie temperature for flavor symmetry breaking in the unstrained case), 
where there is no need to account for non-local diagrams at any temperature. An equally interesting regime is that of non-zero but small, strain, such that $M_f|\mathbf{\epsilon}|\sim \Theta_c$. Mean-field studies showed that the system has an intrinsic tendency toward a nematic state, and therefore a fairly small strain is sufficient in order to drive the ground state towards a strongly renormalized anisotropic semimetal~\cite{parker2021strain}, in which the energy scales are dictated by interaction rather than the symmetry breaking strain field. By including non-local diagrams, our framework allows for controlled calculations in this regime, beyond mean-field theory. Due to the explicit symmetry breaking in this case, we can avoid the divergent susceptibility found in Ref.~\cite{vituri2026controlled}.

\acknowledgements{This work was supported by NSF-BSF Award No. DMR-2310312, the Simons Foundation Collaboration on New Frontiers in Superconductivity (Grant SFI-MPS-NFS-00006741-03), and CRC 183 of the Deutsche Forschungsgemeinschaft (Project C02). 
}

\bibliography{ref}

\begin{widetext}
\appendix
\section{Diagrammatic subtraction rules}
\label{app:diagrams}

As explained in Sec.~\ref{sec:controlled_expansion_C}, when calculating the corrections to the self-energy we subtract the 1PR part of the $n$--point correlation function with respect to the $f$ propagator. Since these subtractions are written using $n$--point correlation functions, the subtraction process is recursive. This gives the alternating minus sign in the expression \eqref{eq:S_def_textual}. As an example of these nested subtractions we look at the following contribution at order $s^6$:
\begin{equation}
\tikzfeynmanset{
  thickarrow/.style={
    /tikz/decoration={
      markings,
      mark=at position 0.5 with {\arrow{Stealth[length=2.4mm, width=1.6mm]}}
    },
    /tikz/postaction=decorate
  },
  doubledashed/.style={
    /tikz/double,
    /tikz/double distance=1.2pt,
    /tikz/dashed,
    thickarrow
  }
}
\centering
\begin{tikzpicture}[scale=0.8, transform shape, baseline=(OL.base)]
  \begin{feynman}

    
    \node at (-1.7,0) {$G_{f,0}^{-1}$};
    \node[draw, regular polygon, regular polygon sides=8, fill=white,
          minimum width=16mm, minimum height=14mm, inner sep=3pt] (OL) at (0,0) {$\Gamma^{(4)}$};
    \draw[doubledashed] (OL.corner 4) to[out=-170, in=-190, looseness=2.5] (OL.corner 3);
    \draw[doubledashed] (OL.corner 2) to[out=100, in=80, looseness=2.5] (OL.corner 1);
    \draw[doubledashed] (OL.corner 8) to[out=10, in=-10, looseness=2.5] (OL.corner 7);
    \node at (1.7,0) {$G_{f,0}^{-1}$};

    \node at (3,0) {$-\,2G_{f,0}^{-1}\,\Biggl($};
    
    \node[draw, regular polygon, regular polygon sides=6, fill=white,
          minimum width=14mm, minimum height=12mm, inner sep=2pt] (HL) at (5.0,0) {$\Gamma^{(3)}$};
    \draw[doubledashed] (HL.corner 3) to[out=-190, in=-235, looseness=2.5] (HL.corner 2);
    \draw[doubledashed] (HL.corner 1) to[out=55, in=10, looseness=2.5] (HL.corner 6);

    \node at (6.3,0) {$-\,$};

    \node[draw, rectangle, rounded corners=1pt, fill=white,
          inner sep=2pt, minimum width=11mm, minimum height=9mm] (S1) at (7.3,0) {$\Gamma^{(2)}$};
    \draw[doubledashed] (S1.north west) to[out=90, in=90, looseness=1.5] (S1.north east);
    
    \node at (8.4,0) {$G_{f,0}^{-1}$};
    
    \node[draw, rectangle, rounded corners=1pt, fill=white,
          inner sep=2pt, minimum width=11mm, minimum height=9mm] (S2) at (9.5,0) {$\Gamma^{(2)}$};
    \draw[doubledashed] (S2.north west) to[out=90, in=90, looseness=1.5] (S2.north east);

    \node at (10.75,0) {$\Biggr) \,\,\,G_{f,0}^{-1}$};
    
    \node[draw, rectangle, rounded corners=1pt, fill=white,
          inner sep=2pt, minimum width=11mm, minimum height=9mm] (S3) at (12.1,0) {$\Gamma^{(2)}$};
    \draw[doubledashed] (S3.north west) to[out=90, in=90, looseness=1.5] (S3.north east);
    
    \node at (13.2,0) {$G_{f,0}^{-1}$};

    \node at (14.3,0) {$-\, \Biggl( G_{f,0}^{-1}$};
    \node[draw, rectangle, rounded corners=1pt, fill=white,
          inner sep=2pt, minimum width=11mm, minimum height=9mm] (S4) at (15.7,0) {$\Gamma^{(2)}$};
    \draw[doubledashed] (S4.north west) to[out=90, in=90, looseness=1.5] (S4.north east);
    
    \node at (17.3,0) {$ \Biggr)^{3} \,G_{f,0}^{-1}\,\, =$};

    \begin{scope}[yshift=-3.0cm, xshift=1.0cm]

    \node at (-1,0) {$G_{f,0}^{-1}$};
    \node[draw, regular polygon, regular polygon sides=8, fill=white,
          minimum width=16mm, minimum height=14mm, inner sep=3pt] (OR) at (0.8,0) {$\Gamma^{(4)}$};
    \draw[doubledashed] (OR.corner 4) to[out=-170, in=-190, looseness=2.5] (OR.corner 3);
    \draw[doubledashed] (OR.corner 2) to[out=100, in=80, looseness=2.5] (OR.corner 1);
    \draw[doubledashed] (OR.corner 8) to[out=10, in=-10, looseness=2.5] (OR.corner 7);
    \node at (2.7,0) {$G_{f,0}^{-1}$};

    \node at (3.7,0) {$-\, 2 G_{f,0}^{-1}$};
    
    \node[draw, regular polygon, regular polygon sides=6, fill=white,
          minimum width=14mm, minimum height=12mm, inner sep=2pt] (HR) at (5.4,0) {$\Gamma^{(3)}$};
    \draw[doubledashed] (HR.corner 3) to[out=-190, in=-235, looseness=2.5] (HR.corner 2);
    \draw[doubledashed] (HR.corner 1) to[out=55, in=10, looseness=2.5] (HR.corner 6);

    \node at (6.8,0) {$G_{f,0}^{-1}$};
    
    \node[draw, rectangle, rounded corners=1pt, fill=white,
          inner sep=2pt, minimum width=11mm, minimum height=9mm] (SR1) at (7.8,0) {$\Gamma^{(2)}$};
    \draw[doubledashed] (SR1.north west) to[out=90, in=90, looseness=1.5] (SR1.north east);
    
    \node at (8.9,0) {$G_{f,0}^{-1}$};

    \node at (9.9,0) {$+\, \Biggl( G_{f,0}^{-1}$};
    \node[draw, rectangle, rounded corners=1pt, fill=white,
          inner sep=2pt, minimum width=11mm, minimum height=9mm] (SR2) at (11.2,0) {$\Gamma^{(2)}$};
    \draw[doubledashed] (SR2.north west) to[out=90, in=90, looseness=1.5] (SR2.north east);
    
    \node at (12.5,0) {$ \Biggr)^{3}\,G_{f,0}^{-1}$};

    \end{scope}

  \end{feynman}
\end{tikzpicture}
\label{eq:AppA1}
\end{equation}
The second term is the subtraction of the 1PI two-loop sub-diagram, where we must recursively subtract the 1PR contribution, times the 1PI one-loop sub-diagram. The factor $2$ is a combinatorial factor that counts the number of distinct ways one can order the sub-diagrams. The third term is the subtraction of three 1PI one-loop sub-diagrams, where there is no extra combinatorial factor, as there is only one distinct ordering for three equivalent sub-diagrams. In general, we must subtract all the different ways one can order sub-diagrams, as they all appear as part of some higher order $n$--point correlation function. For example, consider the following expression
\begin{equation}
\tikzfeynmanset{
  thickarrow/.style={
    /tikz/decoration={
      markings,
      mark=at position 0.5 with {\arrow{Stealth[length=2.4mm, width=1.6mm]}}
    },
    /tikz/postaction=decorate
  },
  doubledashed/.style={
    /tikz/double,
    /tikz/double distance=1.2pt,
    /tikz/dashed,
    thickarrow
  }
}
\centering
\begin{tikzpicture}[scale=0.8, transform shape, baseline=(R1sq.base)]
  \begin{feynman}
    
    \tikzset{
      rect/.style={draw, rectangle, rounded corners=1pt, fill=white,
                   inner sep=2pt, minimum width=12mm, minimum height=10mm}
    }

    \node at (-1,0) {$\Bigg($};
    \node[rect] (R1sq) at (0,0) {$\Gamma^{(2)}$};
    \draw[doubledashed] (R1sq.north west) to[out=90, in=90, looseness=1.2] (R1sq.north east);

    \node at (1.3,0) {$G_{f,0}^{-1}\Bigg)^2$}; 

    \node[rect] (R1a) at (2.6,0) {$\Gamma^{(2)}$}; 
    \node[rect] (R1b) at (5.0,0) {$\Gamma^{(2)}$}; 

    \draw[doubledashed] (R1a.north east) -- (R1b.north west);
    \draw[doubledashed] (R1a.south east) -- (R1b.south west);
    \draw[doubledashed] (R1a.north west) to[out=45, in=135] (R1b.north east);

  \end{feynman}
\end{tikzpicture}
\label{eq:AppA2}
\end{equation}
This term consists of 3 sub-diagrams that can be ordered in three distinct ways. These different orderings are subtracted from 3 different diagrams:
\begin{equation}
\tikzfeynmanset{
  thickarrow/.style={
    /tikz/decoration={
      markings,
      mark=at position 0.5 with {\arrow{Stealth[length=2.4mm, width=1.6mm]}}
    },
    /tikz/postaction=decorate
  },
  doubledashed/.style={
    /tikz/double,
    /tikz/double distance=1.2pt,
    /tikz/dashed,
    thickarrow
  }
}
\centering
\begin{tikzpicture}[scale=0.75, transform shape, baseline=(O2.base)]
  \begin{feynman}
    \tikzset{
      rect/.style={draw, rectangle, rounded corners=1pt, fill=white,
                   inner sep=2pt, minimum width=11mm, minimum height=9mm},
      hex/.style={draw, regular polygon, regular polygon sides=6, fill=white,
                  minimum width=14mm, minimum height=12mm, inner sep=2pt},
      oct/.style={draw, regular polygon, regular polygon sides=8, fill=white,
                  minimum width=16mm, minimum height=14mm, inner sep=2pt},
    }

    \node[oct] (O2) at (0,0) {$\Gamma^{(4)}$};
    \draw[doubledashed] (O2.corner 4) to[out=-170, in=-190, looseness=2.5] (O2.corner 3);
    \draw[doubledashed] (O2.corner 2) to[out=100, in=80, looseness=2.5] (O2.corner 1);

    \node[rect] (R2) at (2.6,0) {$\Gamma^{(2)}$};
    \draw[doubledashed] (O2.corner 7) -- (R2.north west);
    \draw[doubledashed] (O2.corner 8) to[out=45, in=135] (R2.north east);
    \draw[doubledashed] (O2.corner 6) -- (R2.south west);

    \node at (4.0,0) {$,$};

    \begin{scope}[xshift=5.8cm]
      \node[hex] (H3a) at (0,0) {$\Gamma^{(3)}$};
      \node[hex] (H3b) at (2.7,0) {$\Gamma^{(3)}$};

      \draw[doubledashed] (H3a.corner 3) to[out=-190, in=-235, looseness=2.5] (H3a.corner 2);
      \draw[doubledashed] (H3b.corner 1) to[out=55, in=10, looseness=2.5] (H3b.corner 6);

      \draw[doubledashed] (H3a.corner 1) to[out=20, in=160] (H3b.corner 2);
      \draw[doubledashed] (H3a.corner 6) -- (H3b.corner 3);
      \draw[doubledashed] (H3a.corner 5) -- (H3b.corner 4);
    \end{scope}

    \node at (10.2,0) {$,$};

    \begin{scope}[xshift=12.0cm]
      \node[rect] (R4) at (0,0) {$\Gamma^{(2)}$};
      \node[oct] (O4) at (2.6,0) {$\Gamma^{(4)}$};

      \draw[doubledashed] (O4.corner 2) to[out=100, in=80, looseness=2.5] (O4.corner 1);
      \draw[doubledashed] (O4.corner 8) to[out=10, in=-10, looseness=2.5] (O4.corner 7);

      \draw[doubledashed] (R4.north east) -- (O4.corner 4);
      \draw[doubledashed] (R4.south east) -- (O4.corner 5);
      \draw[doubledashed] (R4.north west) to[out=30, in=150] (O4.corner 3);
    \end{scope}

  \end{feynman}
\label{eq:AppA3}
\end{tikzpicture}
\end{equation}
Each diagram in \eqref{eq:AppA3} must come with one subtraction of \eqref{eq:AppA2}. The total combinatorial factor of the subtraction is the number of distinct ways one can order the sub-diagrams of \eqref{eq:AppA2}, which is 3. The sign of the contribution \eqref{eq:AppA2} is the same as the sign of the terms in \eqref{eq:AppA3}, since \eqref{eq:AppA2} contains three sub-diagrams [see the discussion above Eq.~\eqref{eq:D_S_relation}].

\section{Other interactions in the THF action}
\label{app:interactions}

The complete THF Hamiltonian has other interaction terms in addition to the $U$ term considered in the action \eqref{eq:original_action}. These terms are quadratic ($W, J$) and quartic ($V$) in $c$ \cite{song2022magic}. 
To leading order these interactions contribute Hartree-Fock corrections to both the $c$- and $f$-electrons. The $W, J$ interactions give corrections of order $s^0$ to the $c$-propagator and of order $N_f s^2$ to the $f$-propagator, while the $V$ interaction gives a correction of order $N_f s^2$ to the $c$-propagator. All these contributions vanish at charge neutrality. Higher order contributions from these terms arises from the following perturbative diagram:

\begin{equation}
\tikzfeynmanset{
  thinarrow/.style={
    /tikz/decoration={
      markings,
      mark=at position 0.5 with {\arrow{Stealth[length=1.3mm, width=0.8mm]}}
    },
    /tikz/postaction=decorate
  }
}
\centering
\begin{tikzpicture}[scale=0.8, transform shape, baseline=(a2.base)]
  \begin{feynman}

    \vertex (a2) at (0, 0);
    \vertex (b2) at (0.7, 0);     
    \vertex (mid1) at (1.7, 0);   
    \vertex (mid2) at (3.1, 0);   
    \vertex (c2) at (4.1, 0);     
    \vertex (d2) at (4.8, 0);

    \vertex (top1) at (1.7, 0.6);
    \vertex (top2) at (3.1, 0.6);

    \diagram* {
        (a2) -- [dashed, thinarrow] (b2) 
             -- [thinarrow] (c2) -- [dashed, thinarrow] (d2),
        (c2) -- [bend right=50, thinarrow] (b2), 
        (b2) -- [bend right=50, dashed, thinarrow] (c2) 
    };

    \node [below=4pt] at (b2) {\footnotesize $J$};
    \node [below=4pt] at (c2) {\footnotesize $J$};


  \end{feynman}
\end{tikzpicture}
\end{equation}
as well as similar diagrams with additional $U$ insertions, and similar diagrams with $W$ replacing $J$. In this diagram, $f$ and $c$ propagators are marked with solid and dashed lines, respectively. These diagrams give a correction of order $s^2$ only for the $c$-electrons. Since the Mott bands are $f$-like at order $s^0$, the correction to the Mott band propagator is of higher order than $s^2$ and we ignore it in our calculation. This correction will affect the dispersion close to the $\gamma_M$ point, and it was taken into account in Ref.~\cite{hu2026twisted}. Similarly, diagrams can be drawn that give $s^2$ corrections to the $f$-$c$ hybridization. The corresponding correction to the Mott bands is again of higher order in $s^2$.

In addition to these interactions, there is a nearest-neighbor interaction term for the $f$-electrons ($U_2)$ and a cubic in $c$ interaction ($K)$~\cite{song2022magic}. These terms are much smaller in magnitude compared to the other interaction terms, and we ignore them.

\section{Single-particle Hamiltonian in the orbital sector basis}
\label{app:symmmetry}

In this appendix, we provide the explicit form of the single-particle Hamiltonian in the orbital basis (in which the generator of the approximate symmetry of Sec.~\ref{sec:strained_symmetry} is diagonal). 
For simplicity we work in the valley $\eta=1$ and take $\epsilon_{xy}=0$. Define

\begin{equation}
    \mathcal{U}(\mathbf k) = \text{diag}\left(U_y, \mathcal B(\mathbf k) U_y, U_y\right)
\end{equation}
with $U_y=\frac{1}{\sqrt{2}}(\mathbbm{1}-i\sigma_x)$ the rotation to the $\sigma_y$ eigenbasis. It is easy to see that 

\begin{equation}
\mathcal U(\mathbf k) \mathcal O (\mathbf k)\mathcal U^\dagger(\mathbf k) = \text{diag}(1,-1,1,-1,1,-1).
\end{equation}
We reorder the resulting basis into the $+$ and $-$ sectors, writing the annihilation operator as $\Phi_k \equiv (c_{+,1}, c_{+,2}, f_+, c_{-,1}, c_{-,2}, f_-)$. The single-particle Hamiltonian in this new basis is then:

\begin{align}
\tilde h_{\mathrm{flat-chiral}}(\mathbf k)
\;\equiv\; \mathcal{\tilde U^\dagger}(\mathbf k)\,h_{\mathrm{flat-chiral}}(\mathbf k)\,\mathcal{\tilde U}(\mathbf k)
&=
\begin{pmatrix}
h_+(\mathbf k) & 0\\
0 & h_-(\mathbf k)
\end{pmatrix},
\end{align}
where $\mathcal{\tilde U}(\mathbf k)=\mathcal{U}(\mathbf k)P$ with $P$ the permutation to the new basis $\Phi_k$, and
\begin{align}
h_+(\mathbf k) &=
\begin{pmatrix}
0 & v_*|\mathbf k| & \gamma\\
v_*|\mathbf k| & 0 & 0\\
\gamma & 0 & +M_f \epsilon_-
\end{pmatrix},
\qquad
h_-(\mathbf k) =
\begin{pmatrix}
0 & v_*|\mathbf k| & \gamma\\
v_*|\mathbf k| & 0 & 0\\
\gamma & 0 & -M_f \epsilon_-
\end{pmatrix}.
\end{align}

Away from the flat-chiral limit, the Hamiltonian in this basis takes the form:

\begin{equation}
\begin{aligned}
\tilde h(\mathbf k)
=
\begin{pmatrix}
0 & v_\star |\mathbf{k}| & \gamma - k_y\,v'_\star & 0 & 0 & k_x\,v'_\star \\[6pt]
v_\star |\mathbf{k}| &
\dfrac{2M k_x k_y}{|\mathbf{k}|^2} &
\dfrac{c''\,\epsilon_-\,k_y}{|\mathbf{k}|} &
0 &
\dfrac{M\,(k_x^2-k_y^2)}{|\mathbf{k}|^2} &
-\dfrac{c''\,\epsilon_-\,k_x}{|\mathbf{k}|} \\[12pt]
\gamma - k_y\,v'_\star &
\dfrac{c''\,\epsilon_-\,k_y}{|\mathbf{k}|} &
- M_f\,\epsilon_- &
k_x\,v'_\star &
\dfrac{c''\,\epsilon_-\,k_x}{|\mathbf{k}|} &
0 \\[10pt]
0 & 0 & k_x\,v'_\star & 0 & v_\star |\mathbf{k}| & \gamma + k_y\,v'_\star \\[6pt]
0 &
\dfrac{M\,(k_x^2-k_y^2)}{|\mathbf{k}|^2} &
\dfrac{c''\,\epsilon_-\,k_x}{|\mathbf{k}|} &
v_\star |\mathbf{k}| &
-\dfrac{2M k_x k_y}{|\mathbf{k}|^2} &
\dfrac{c''\,\epsilon_-\,k_y}{|\mathbf{k}|} \\[12pt]
k_x\,v'_\star &
-\dfrac{c''\,\epsilon_-\,k_x}{|\mathbf{k}|} &
0 &
\gamma + k_y\,v'_\star &
\dfrac{c''\,\epsilon_-\,k_y}{|\mathbf{k}|} &
M_f\,\epsilon_-
\end{pmatrix}.
\end{aligned}
\end{equation}
We use this form of the Hamiltonian in Fig.~\ref{fig:ideal}.

\section{Strained MATBG spectral features at tree-level}
\label{app:tree_level}

As explained briefly in Sec.~\ref{sec:strained_tree_level}, if we consider the single-site $f$ Hamiltonian:

\begin{equation}
    H =M_f |\mathbf \epsilon|\sum_i \left(f^\dagger_{i+} f_{i+} - f^\dagger_{i-} f_{i-}\right)
+ \frac{U}{2}\,(n_{f+} + n_{f-} - 4)^2 \,
\end{equation}
then at charge neutrality we have a full $f_-$ band and an empty $f_+$ band. Any diagrammatic process we describe in imaginary time will vanish if it includes the propagation of a higher-energy band $f_+$ hole or a lower-energy band $f_-$ electron, due to Pauli blocking. 
As a result, the only non-vanishing diagram is the Hartree-Fock correction, and that is the full effect of $U$ to tree-level in the hybridization. This also explains the appearance of two Dirac cones away from the $\gamma$ and $\kappa, \kappa'$ points, as we know that the non-interacting strained Hamiltonian has two Dirac cones, and that the strain unpins these Dirac cones from the corners of the mini Brillouin zone (BZ).

Lastly, this simple single-particle picture for the strained Hamiltonian at tree-level allows us to perform explicit calculations of different quantities. For example, we consider the shift of the Mott bands as a result of the hybridization. At large $k$ this can be done by 2nd order perturbation theory:

\begin{equation}
    H_{eff}=H_{ff}-H_{fc}\frac{1}{E-H_{cc}}H_{cf}.
\end{equation}
Inserting the expressions from \eqref{eq:thf_hamiltonian} one obtains:

\begin{equation}
    \delta E_\pm(\hat k, |k|) \approx \frac{u+M_f|\mathbf \epsilon|}{v_\star^2|k|}\,
    2\gamma v_\star' (\hat k^{(\eta)} \cdot \hat \epsilon^{(\eta)}),
\end{equation}
where $\pm$ refers to the upper and lower energy Mott bands, respectively, $\hat \epsilon^{(\eta)}$ is the direction of the strain vector as defined in \ref{sec:strained_ham}, and $\hat{k}^{(\eta)}$ the direction of the momentum. This is a shift of the Mott bands that depends on the relative angle between $\hat k^{(\eta)}, \hat \epsilon^{(\eta)}$. For standard model parameters like the ones given in Table \ref{table:params}, this shift is of order $\sim 1\,\text{meV}$.

\section{One-loop self-energy}
\label{app:one_loop}

We start with Eq.~\eqref{eq:one_loop_diag_expression} of the main text:
\begin{equation}
\begin{split}
\tilde\gamma^\dagger G_{f,0} S^{(1)}_\lambda(i\omega) G_{f,0} \tilde\gamma
=
\sum_{\lambda'}
\int_{A_{BZ}} \frac{d^2k'}{A_{BZ}} \int_{-\infty}^{\infty}\frac{d\omega'}{2\pi}\; 
 G^{0}_{c,\lambda';a_1 a_2}(\mathbf{k}',i\omega')\,
\Gamma^{(2)}_{\lambda, \lambda'}(\omega,\omega', \omega', \omega)\,,
\end{split}
\label{eq:app_S1}
\end{equation}
where we defined the rotated $\tilde\gamma=U^\dagger\hat\gamma$, with $U$ the rotation to the strain eigenbasis. Using the four–point function~\eqref{eq:4_point_func_result} in this basis with $\omega_1 = \omega'_1 = \omega$ and $\omega_2 = \omega'_2 = \omega'$
we obtain
\begin{align}
\Big\langle
f_{\lambda,b}(i\omega)\,
f_{\lambda',b_1}(i\omega')\,
\bar f_{\lambda',b_1}(i\omega')\,
\bar f_{\lambda,b}(i\omega)
\Big\rangle_{c,0}
&=
\,(1-\delta_{b,b_1})
\Big[
\delta_{b,+}\,\bigg(\frac{1}{(u+M_f|\mathbf \epsilon|)-i\omega}+
\frac{1}{(u+M_f|\mathbf \epsilon|)+i\omega'}\bigg)^2
\frac{1}{2M_f|\mathbf \epsilon| - i(\omega-\omega')}
\nonumber\\
&\hspace{1.5cm}
+\,\delta_{b,-}\,\bigg(\frac{1}{(u+M_f|\mathbf \epsilon|)-i\omega'}+
\frac{1}{(u+M_f|\mathbf \epsilon|)+i\omega}\bigg)^2
\frac{1}{2M_f|\mathbf \epsilon| - i(\omega'-\omega)}
\Big].
\end{align}
One can see that Eq.~\eqref{eq:app_S1} is therefore diagonal in the strain eigenbasis. We find
\begin{equation}
    S^{(1)}(i\omega) = 
    \begin{pmatrix}
        S_+(i \omega) & 0 \\
        0 & S_-(i \omega)
    \end{pmatrix}
\end{equation}
with
\begin{equation}
\begin{split}
S^{(1)}_+(i\omega)
=
\int_{-\infty}^{\infty}\frac{d\omega'}{2\pi}\;
\mathcal{ I}_{-}(i\omega')
\left[1+\frac{(u+M_f|\mathbf \epsilon|)-i\omega}{(u+M_f|\mathbf \epsilon|)+i\omega'}\right]^2 
 \frac{1}{2M_f|\mathbf \epsilon| - i(\omega-\omega')},
\end{split}
\end{equation}

\begin{equation}
\begin{split}
S^{(1)}_-(i\omega)
=
\int_{-\infty}^{\infty}\frac{d\omega'}{2\pi}\;
\mathcal{ I}_+(i\omega')
\left[1+\frac{(u+M_f|\mathbf \epsilon|)+i\omega}{(u+M_f|\mathbf \epsilon|)-i\omega'}\right]^2
\frac{1}{2M_f|\mathbf \epsilon| - i(\omega'-\omega)},
\end{split}
\end{equation}
where $\mathcal{I}_\pm$ is defined below Eq.~\eqref{eq:S_pm}, and we used:

\begin{align}
\sum_{\lambda'} \int_{A_{BZ}} \frac{d^2k'}{A_{BZ}}\;
(\tilde\gamma^\dagger)_{a_2 b_1}\,
G^{0}_{c,\lambda';a_1 a_2}(\mathbf{k}',i\omega')\,
\tilde\gamma_{b_1 a_1}
&=
\sum_{\lambda'} \int_{A_{BZ}} \frac{d^2k'}{A_{BZ}}\;
(\hat\gamma^\dagger U)_{a_2 b_1}\,
G^{0}_{c,\lambda';a_1 a_2}(\mathbf{k}',i\omega')\,
(U^\dagger\hat\gamma)_{b_1 a_1}
\nonumber\\
&=
U^\dagger_{b_1 b_2}\,
\mathcal I(i\omega')_{b_2 b_3}\,
U_{b_3 b_1}
=
{\mathcal I}_{b_1}(i\omega').
\end{align}

To get an explicit expression for $\operatorname{Im} S^{(1)}_+(\omega)$ we express $\mathcal{ I}_\pm$ using the spectral representation, where we define 
\begin{equation}
    \rho_\pm(\omega)
=
-\frac{1}{\pi}\,\rm{Im}  \mathcal I_\pm(\omega + i0^+).
\end{equation}
We have
\begin{equation}
S^{(1)}_+(i\omega)
=
\int_{-\infty}^{\infty}\frac{d\omega'}{2\pi}
\int_{-\infty}^{\infty} d\varepsilon\;
\frac{{\rho}_-(\varepsilon)}{i\omega'-\varepsilon}
\left[
1+\frac{(u+M_f|\epsilon|)-i\omega}
        {(u+M_f|\epsilon|)+i\omega'}
\right]^2
\frac{1}{2M_f|\epsilon|-i(\omega-\omega')}.
\label{eq:Splus_spectral_start}
\end{equation}
We perform the $\omega$ integral using contour integration, closing the contour with a semi-circle below the real axis. The result is nontrivial only for $\varepsilon>0$:
\begin{equation}
S^{(1)}_+(i\omega)
=
-\int_{0}^{\infty} d\varepsilon\;
{\rho}_-(\varepsilon)
\left[
1+\frac{(u+M_f|\epsilon|)-i\omega}
        {(u+M_f|\epsilon|)+\varepsilon}
\right]^2
\frac{1}{2M_f|\epsilon|-i(\omega-\varepsilon)} .
\end{equation}
By performing a Wick rotation we find
\begin{equation}
    \textrm{Im} S_+^{(1)}(\omega) = -\pi\theta(\omega - 2M_f|\mathbf \epsilon|)
    \rho_-(\omega-2M_f|\mathbf \epsilon|)
    \left[\frac{2u}{u-M_f|\mathbf \epsilon|+\omega}
    \right]^2,
\end{equation}
which is Eq.~\eqref{eq:S_imag} of the main text for the $+$ sector. The equation for the $-$ sector is obtained similarly.

$\mathcal{I}$ can be calculated explicitly in the flat-chiral limit. Working in the chiral basis we have:

\begin{equation}
G_{f,0}(i\omega;\phi)
=
\frac{
 i\omega\,\sigma_0
 - (u+M_f)\bigl[\sin(2\phi)\,\sigma_x + \cos(2\phi)\,\sigma_y\bigr]
}{
 (i\omega)^2 - (u+M_f)^2
}.
\end{equation}
With $\phi$ the heterostrain angle. This yields:
\begin{equation}
G_{c,0}^{-1}(\mathbf{k},i\omega)
=
i\omega\,\sigma_0\otimes\zeta_0
-\Bigl[
v_\ast k_x\,\sigma_0\otimes\zeta_x
- v_\ast k_y\,\sigma_z\otimes\zeta_y
+ \Sigma_c(i\omega;\phi)\otimes\frac{\zeta_0+\zeta_z}{2}
\Bigr],
\end{equation}
where $\zeta_{x,y,z}$ are Pauli matrices, with $\zeta_{z}$ having an eigenvalue $+1$ for the $\Gamma_3$ $c$-electron block and eigenvalue $-1$ for the $\Gamma_1\oplus\Gamma_2$ block. In the flat-chiral limit only the $\Gamma_3$ block of $G_{c,0}$ is needed for the calculation of $\mathcal{ I}$ since only the $\Gamma_3$ $c$-electrons hybridize with the $f$-electrons. For that block we have:

\begin{equation}
\small
G^{(\Gamma_3)}_{c,0}(\mathbf{k},i\omega;\phi)
=
\frac{
 i\omega\bigl[(i\omega)^2-\Delta^2\bigr]
 \Bigl[((i\omega)^2-v_*^2k^2)\bigl((i\omega)^2-\Delta^2\bigr)-\gamma^2(i\omega)^2\Bigr]\sigma_0
 - \gamma^2(i\omega)^2\Delta\bigl[(i\omega)^2-\Delta^2\bigr]
 \Bigl[\sin(2\phi)\sigma_x+\cos(2\phi)\sigma_y\Bigr]
}{
\Bigl[((i\omega)^2-v_*^2k^2)\bigl((i\omega)^2-\Delta^2\bigr)-\gamma^2(i\omega)^2\Bigr]^2
-\gamma^4\Delta^2(i\omega)^2
},
\end{equation}
where we define $\Delta=u+M_f|\mathbf \epsilon|$.

We start by calculating the diagonal part of $\mathcal I$, given by the $\sigma_0$ part of $G^{(\Gamma_3)}_{c,0}(\mathbf{k},i\omega;\phi)$:
\begin{equation}
\begin{aligned}
\mathcal I_{b,b}(i\omega)
&=
\frac{2\pi i\omega\, \tilde N_f \gamma^2}{A_{BZ}}
\int_0^{\sqrt{A_{BZ}/\pi}} dk\,k\;
\frac{
 \bigl[(i\omega)^2-\Delta^2\bigr]
 \Bigl[((i\omega)^2-v_*^2k^2)\bigl((i\omega)^2-\Delta^2\bigr)-\gamma^2(i\omega)^2\Bigr]
}{
 \Bigl[((i\omega)^2-v_*^2k^2)\bigl((i\omega)^2-\Delta^2\bigr)-\gamma^2(i\omega)^2\Bigr]^2
 - \gamma^4 \Delta^2 (i\omega)^2
}
\\[6pt]
&=
-\frac{\pi i\omega\, \tilde N_f \gamma^2}{2 v_*^2 A_{BZ}}
\ln\Bigl(
\bigl[((i\omega)^2-v_*^2k^2)\bigl((i\omega)^2-\Delta^2\bigr)-\gamma^2(i\omega)^2\bigr]^2
-\gamma^4 \Delta^2 (i\omega)^2
\Bigr)
\Big|_{k=0}^{k=\sqrt{A_{BZ}/\pi}}
\\[6pt]
&=
- i\omega \frac{\tilde N_f}{2}s^2
\ln\Biggl(
 \frac{
 \Bigl[((i\omega)^2-\frac{\gamma^2}{s^2})((i\omega)^2-\Delta^2)-\gamma^2(i\omega)^2\Bigr]^2
 - \gamma^4 \Delta^2 (i\omega)^2
 }{
 \Bigl[(i\omega)^2((i\omega)^2-\Delta^2)-\gamma^2(i\omega)^2\Bigr]^2
 - \gamma^4 \Delta^2 (i\omega)^2
 }
\Biggr)
\\[6pt]
&=
- i\omega \frac{\tilde N_f}{2}s^2
\ln\Biggl(
1
- \frac{2\gamma^2((i\omega)^2-\Delta^2)}{s^2}
\frac{(i\omega)^2((i\omega)^2-\Delta^2)-\gamma^2(i\omega)^2}{
\bigl[(i\omega)^2((i\omega)^2-\Delta^2)-\gamma^2(i\omega)^2\bigr]^2
-\gamma^4\Delta^2(i\omega)^2}
\\
&\qquad\qquad
+
\frac{\gamma^4((i\omega)^2-\Delta^2)^2}{
s^4\Big(\bigl[(i\omega)^2((i\omega)^2-\Delta^2)-\gamma^2(i\omega)^2\bigr]^2
-\gamma^4\Delta^2(i\omega)^2\Big)}
\Biggr)
\\[6pt]
&\approx
- i\omega \frac{\tilde N_f}{2}s^2
\ln\Biggl(
1 + \frac{2}{s^2} + \frac{( (i\omega)^2-\Delta^2 )}{s^4 (i\omega)^2}
\Biggr).
\end{aligned}
\end{equation}
In the approximation we used $v_\star^2A_{BZ}\gg \gamma^2 \gg \Delta \sim \omega$. We also defined $\tilde N_f=\frac{N_f}{2}$ the spin and valley degeneracy. Wick rotating, the argument of the log is negative, yielding:
\begin{equation}
\textrm{Im}\,\mathcal I_{bb}^R(\omega) \;\simeq\; -\,\pi\,|\omega|\,\frac{\tilde N_f}{2} s^2
\end{equation}

For the off-diagonal part, we calculate the prefactor of $\Bigl[\sin(2\phi)\sigma_x+\cos(2\phi)\sigma_y\Bigr]$. We have:
\begin{equation}
\begin{aligned}
&-\frac{2\pi \tilde N_f \gamma^4 (i\omega)^2 \Delta}{A_{BZ}}
\int_0^{\sqrt{A_{BZ}/\pi}} dk\,k\;
\frac{( (i\omega)^2-\Delta^2 )}
{
\Bigl[((i\omega)^2-v_*^2k^2)((i\omega)^2-\Delta^2)-\gamma^2(i\omega)^2\Bigr]^2
- \gamma^4 \Delta^2 (i\omega)^2
}
\\[6pt]
&=
-\frac{\pi \tilde N_f \gamma^4 (i\omega)^2 \Delta}{A_{BZ}}
\int_0^{A_{BZ}/\pi} dt\;
\frac{( (i\omega)^2-\Delta^2 )}
{
\Bigl[((i\omega)^2-v_*^2t)((i\omega)^2-\Delta^2)-\gamma^2(i\omega)^2\Bigr]^2
- \gamma^4 \Delta^2 (i\omega)^2
}
\\[6pt]
&\stackrel{\substack{
y=((i\omega)^2-v_*^2 t)((i\omega)^2-\Delta^2)-\gamma^2(i\omega)^2\\
dy=-( (i\omega)^2-\Delta^2 )v_*^2\,dt
}}{=}
\frac{\pi \tilde N_f \gamma^4 (i\omega)^2 \Delta}{v_*^2 A_{BZ}}
\int_{y(0)}^{y(A_{BZ}/\pi)} dy\;
\frac{1}{y^2-\gamma^4 \Delta^2 (i\omega)^2}
\\[6pt]
&=
\frac{\pi \tilde N_f \gamma^2 (i\omega)}{2 v_*^2 A_{BZ}}
\ln\!\left(
\frac{y-\gamma^2 \Delta (i\omega)}{y+\gamma^2 \Delta (i\omega)}
\right)
\Bigg|_{y(0)}^{y(A_{BZ}/\pi)}
\\[6pt]
&=
\frac{\tilde N_f}{2} (i\omega)s^2
\ln\!\left(
\frac{y(A_{BZ}/\pi)-\gamma^2 \Delta (i\omega)}{y(A_{BZ}/\pi)+\gamma^2 \Delta (i\omega)}
\cdot
\frac{y(0)+\gamma^2 \Delta (i\omega)}{y(0)-\gamma^2 \Delta (i\omega)}
\right)
\\[6pt]
&\approx
\frac{\tilde N_f}{2} (i\omega)s^2
\ln\!\left(
\left(1+\frac{1}{s^2}\frac{i\omega-\Delta}{i\omega}\right)
\left(1+\frac{1}{s^2}\frac{i\omega+\Delta}{i\omega}\right)^{-1}
\right)
\\[6pt]
&\approx
\frac{\tilde N_f}{2} (i\omega)s^2
\ln\!\left(\frac{i\omega-\Delta}{i\omega+\Delta}\right)
\;\xrightarrow{i\omega\rightarrow \omega+i0^+}\;
\frac{\tilde N_f}{2}\omega s^2
\ln\!\left(
\frac{\omega+i0^+-\Delta}{\omega+i0^++\Delta}
\right).
\end{aligned}
\end{equation}
Since the Pauli matrices are Hermitian, the anti-Hermitian part of $\mathcal I$ is the imaginary part of this prefactor. We thus get that the anti-Hermitian part is:
\begin{equation}
    \pi \omega \frac{\tilde N_f}{2} s^2
    \,\bigl[\sin(2\phi)\,\sigma_x + \cos(2\phi)\,\sigma_y\bigr].
\end{equation}
Combining the results and rotating to the strain eigenbasis, we have:
\begin{equation}
     \rho_\pm(\omega) = - \tilde N_f s^2 |\omega| \theta(\pm\omega),
\end{equation}
which scales as $\tilde N_f s^2 \omega$ as stated in the main text. The Heaviside functions result in $\operatorname{Im} S^{(1)}$ vanishing, as expected for the flat-chiral limit.

\section{Additional spectral function results}
\label{app:extra}
We present here supplementary results for the one-loop spectral function calculations. 
\begin{figure*}
    \centering
    \includegraphics[width=\linewidth]{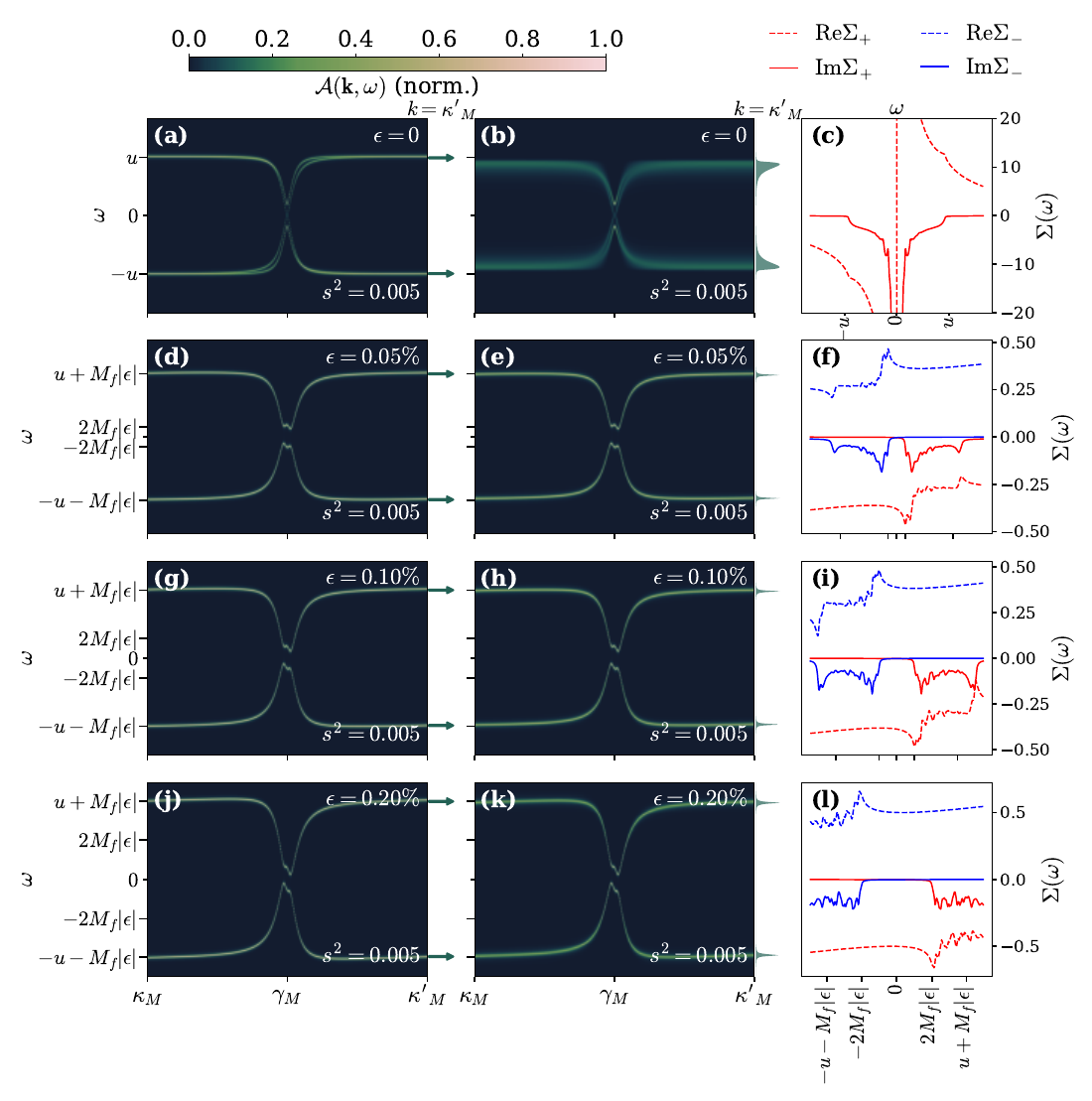}
    \caption{Same as Fig.~\ref{fig:spectrum} except that the same value for $v_\star$ is used for both the strained and unstrained cases. (a) Spectral function $\mathcal{A}(\mathbf k,\omega)$ along the $\kappa_M$-$\gamma_M$-$\kappa'_M$ line calculated to tree-level for the unstrained case. (b) Unstrained spectral function to one-loop order, (c) the self-energy to one-loop order. To the side of panels (a), (b) we show the spectral cut $\mathcal{A}(\kappa'_M,\omega)$, where for (a) the delta functions of the tree-level spectrum are indicated with arrows. Both calculations are made as explained in Sec.~\ref{sec:controlled_expansion_C} using Eq.~\eqref{eq:S_definition}, \eqref{eq:f_self_energy}, \eqref{eq:c_self_energy}. Panels (d)-(f), (g)-(i), (j)-(l) show the same results for the cases of $0.05\%$, $0.1\%$, $0.2\%$ strain, respectively. For the unstrained case there is no high or low energy sector and thus $\Sigma=\Sigma_+=\Sigma_-$. Apart from the strain magnitude, all parameters are the same for all graphs.}
    \label{fig:app_spec}
\end{figure*}
Fig.~\ref{fig:app_spec} shows the difference in the Mott bands' width for the strained compared to the unstrained case for the same parameters. As can be seen, for the same set of parameters the Mott bands in the strained case are sharp, while they acquire significant width in the unstrained case.

\begin{figure}
    \centering
    \includegraphics[width=\linewidth]{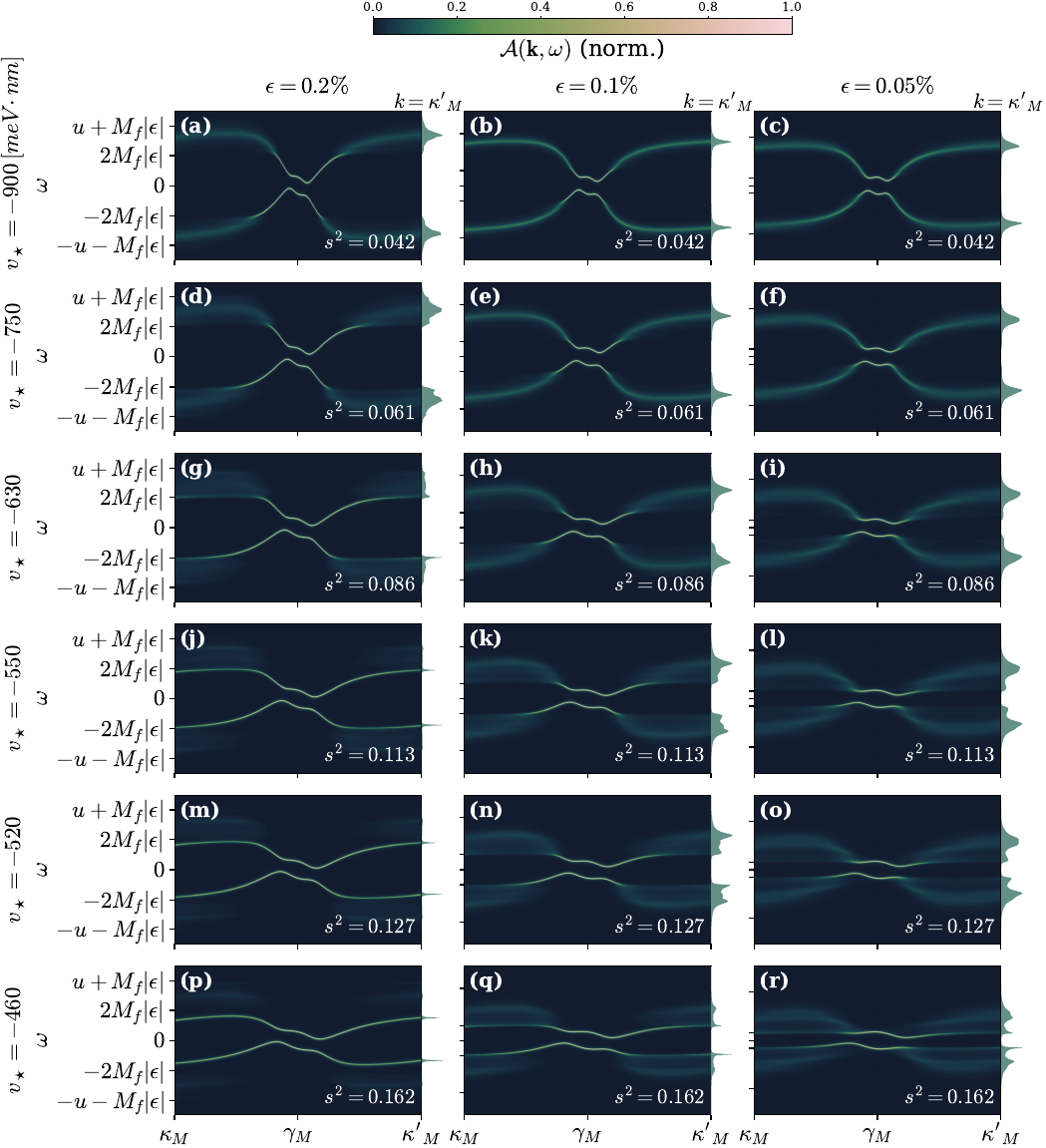}
    \caption{Spectral function computed to one-loop order from Eq.~\eqref{eq:S_definition}, \eqref{eq:f_self_energy}, \eqref{eq:c_self_energy}. To the right of each panel we plot the spectral cut $\mathcal{A}(\kappa'_M,\omega)$, normalized relative to its maximal value. From left to right the strain takes the values $0.02\%, 0.01\%, 0.05\%$, and from top to bottom $v_\star$ takes the values $-900, -750, -630, -550, -520, -460$. All other parameters are the same in all plots.}
    \label{fig:app_grid}
\end{figure}

Fig.~\ref{fig:app_grid} shows the strained spectral function to one-loop order for different strain magnitudes and $v_\star$ values. Lower values of $v_\star$ result in a larger $s^2$. As explained in the main text, we see the spectral weight moving from the Mott band to the trion band as $s^2$ gets larger. For smaller values of $s^2$, the spectral weight on the trion band is smaller, and at some point the feature appears as a ``kink" in the spectrum similar to the one known from the electron-phonon problem, with the troin band at $2M_f|\mathbf \epsilon|$ barely visible. As seen in Fig.~\ref{fig:app_grid}, the exact value of $s^2$ where this crossover occurs depends on the strain magnitude. 

\section{Holstein electron-phonon toy model}
\label{app:hosltein}

In strained MATBG, the trion found in this paper is the result of the coupling between single-particle excitations and a finite energy bosonic mode (an $f$-exciton). In order to understand this effect, we calculate the electron's self-energy to one-loop order for the Holstein model, the prototypical example of coupling to a finite energy bosonic mode, following Ref.~\cite{marsiglio2008electron}. Our starting point is the Holstein model,
\begin{equation}
H =
\sum_{k\sigma} \epsilon_k c^\dagger_{k\sigma} c_{k\sigma}
+ \sum_q \hbar \omega_E a_q^\dagger a_q
+ \frac{g}{\sqrt{N}} \sum_{kq\sigma}
\left( a_q + a^\dagger_{-q} \right)
c^\dagger_{k+q,\sigma} c_{k\sigma},
\end{equation}
where $\omega_E$ is the Einstein phonon frequency, $\epsilon_k$ the band dispersion, $g$ the electron-phonon coupling constant and $N$ the system size. The operator $c_{k\sigma}$ is the annihilation operator for an electron with momentum $k$ and spin $\sigma$, while $a_q$ is the annihilation operator for a phonon with momentum $q$. Using the approximation of a constant DOS and an infinite bandwidth, the one-loop self-energy for the electrons is given by:
\begin{equation}
\Sigma(i\omega_m)
=
\lambda \omega_E^2
\int_{-\infty}^{\infty} d\epsilon \,
\frac{1}{\beta}
\sum_{m'}
\frac{1}{\omega_E^2 + (\omega_{m'}-\omega_m)^2}
\frac{1}{i\omega_{m'} - (\epsilon-\mu)} ,
\end{equation}
with $\lambda= \frac{2g^2 N(\epsilon_F)}{\omega_E}$ the dimensionless electron-phonon coupling, $N(\epsilon_F)$ the constant DOS and $\beta$ the inverse temperature. The sum over $m'$ is a sum over fermionic Matsubara frequencies. Using standard Matsubara summation we can perform the integral and obtain the self-energy at zero temperature:
\begin{equation}
\Sigma(\omega + i0^+) =
\lambda \omega_E
\left[
\ln \left| \frac{\omega_E - \omega}{\omega_E + \omega} \right|
- i \pi \, \theta(|\omega| - \omega_E)
\right],
\end{equation}
As given in Eq.~\eqref{eq:holstein_self_energy} in the main text. 
The Heaviside function is a consequence of the threshold for electron decay set by the finite energy of the phonons. This Heaviside function in the imaginary part is accompanied by a logarithmic divergence of the real part, as a result of the Kramers-Kronig relations.

Setting the chemical
potential $\mu = 0$, the poles of the electron's Green's function are determined by
\begin{equation}
\omega = \epsilon_k + \rm{Re} \Sigma(\omega) .
\end{equation}
For $|\omega| \gg \omega_E$ the real part of the self-energy vanishes,
and the dispersion is approximately given by $\epsilon_k$.
Below $\omega_E$ the equation for the pole's location becomes
\begin{equation}
\omega =
\epsilon_k +
\lambda \omega_E
\ln \left| \frac{\omega_E - \omega}{\omega_E + \omega} \right| .
\end{equation}
This equation always admits a solution $\omega_t < \omega_E$, so long as $\epsilon_k>\omega_E$.
For large $\epsilon_k$ the solution approaches $\omega_E$ from below,
corresponding to a phonon-induced satellite feature.
The residue of this excitation is given by
\begin{equation}
Z =
\left[
1 - \left. \frac{d\, \rm{Re} \Sigma(\omega)}{d\omega}
\right|_{\omega = \omega_t}
\right]^{-1} .
\end{equation}
Solving for large $\epsilon_k$, equivalently for a pole just below the phonon energy,
\begin{equation}
\omega_t = \omega_E(1-\eta), \qquad 0<\eta \ll 1,
\end{equation}
to first order one obtains $\eta \simeq 2 \exp\!\left(\frac{1-\epsilon_k/\omega_E}{\lambda}\right)$ and therefore:
\begin{equation}
\omega_t \simeq \omega_E\left[1-2\exp\!\left(\frac{1-\epsilon_k/\omega_E}{\lambda}\right)\right]
.
\end{equation}
To first order in $\eta$, the residue of the excitation at $\omega_t$ is given by
\begin{equation}
Z \simeq \left(1+\frac{\lambda}{\eta}\right)^{-1}
= \frac{\eta}{\lambda} + O(\eta^2).
\end{equation}
Using the result for $\eta$ we get
\begin{equation}
Z \simeq
\frac{2}{\lambda}\exp\!\left(\frac{1-\epsilon_k/\omega_E}{\lambda}\right),
\end{equation}
which vanishes exponentially in $\frac{\epsilon_k}{\omega_E}$.

\section{QTM spectrum of the BM model}
\label{app:bm_qtm_spectrum}

We briefly present the Bistritzer-MacDonald (BM) model for unstrained and strained TBG, and show the QTM spectrum as calculated from Eq.~\eqref{eq:QTM_spectrum} for the non-interacting BM model.

\subsection{BM Hamiltonian}
\label{app:BM_ham}

The BM Hamiltonian for the $K$ valley of TBG is given by \cite{bistritzer2011moire}:
\begin{equation}
\hat{H}^0_{ij}(\mathbf{k}) =
\begin{pmatrix}
h_{\mathbf{k}+g_i}\!\left(\tfrac{\theta_{\rm TBG}}{2}\right)\delta_{ij} 
& \displaystyle\sum_{n=1}^{3}\hat{T}_n\,\delta_{g_i+g_n,g_j}\\[6pt]
\displaystyle\sum_{n=1}^{3}\hat{T}_n^\dagger\,\delta_{g_i,g_j+g_n}
& h_{\mathbf{k}+g_i}\!\left(-\tfrac{\theta_{\rm TBG}}{2}\right)\delta_{ij}
\end{pmatrix},
\label{eq:BM_ham}
\end{equation}
with moir\'e reciprocal lattice vectors $g_1 = 0$ and $g_{2,3} = k_M(\pm\sqrt{3}/2,\, 3/2)^T$, where $k_M = 2|K|\sin(\theta_\mathrm{TBG}/2)$ and $\theta_{TBG}=1.05^\circ$ the twist angle. 
The single-layer Dirac Hamiltonian for a layer rotated by $\theta$ is
\begin{equation}
h_{\mathbf{p}}(\theta) = \hbar v_D\bigl(O(\theta)^T\mathbf{p} - K\bigr)\cdot\boldsymbol{\sigma},
\end{equation}
with $O(\theta)$ the clockwise rotation matrix, $v_D$ the bare Dirac velocity, and $\boldsymbol\sigma = (\sigma_x, \sigma_y)$ Pauli matrices in the sublattice basis. 
The interlayer tunneling matrices are
\begin{equation}
\hat{T}_n = w_0\,\mathbb{I} 
+ w_1\!\left(\cos\tfrac{2\pi(n-1)}{3}\,\sigma_x + \sin\tfrac{2\pi(n-1)}{3}\,\sigma_y\right),
\end{equation}
where $w_0, w_1$ are the intra-sublattice and inter-sublattice hopping magnitudes, respectively.

\subsection{Effect of strain}
\label{app:BM_strain}

Uniform uniaxial heterostrain is parametrized by a magnitude $\epsilon$ and direction $\phi$, with the strain tensor
\begin{equation}
\mathcal{E}_l = O(\phi)^T
\begin{pmatrix}
\epsilon & 0\\0 & -\nu_P\epsilon
\end{pmatrix}
O(\phi)=
\begin{pmatrix}
    \epsilon_{xx} & \epsilon_{xy} \\
    \epsilon_{yx} & \epsilon_{yy}
\end{pmatrix}
,
\end{equation}
where $\nu_P\approx 0.16$ is the Poisson ratio of graphene. We take $\mathcal{E}_t=-\mathcal{E}_{b}=\mathcal{E}/2$. With the addition of strain, the single-layer Hamiltonian in the $K$ valley becomes~\cite{bi2019designing, wei2025dirac}
\begin{equation}
h'_{\mathbf{p},l} = \hbar v_D\bigl(M_l^T\mathbf{p} - K - A_l\bigr)
\cdot(\sigma_x,\,\sigma_y),
\label{eq:strained_dirac}
\end{equation}
where $M_l = (1+\mathcal{E}_l)O(\theta_l)$, and $A_t = -A_b = \frac{\sqrt{3}\,\beta}{4a_0}\bigl(\epsilon_{xx}-\epsilon_{yy},\;-2\epsilon_{xy}\bigr)$, $a_0$ the graphene lattice constant and $\beta\approx 3.14$. The strained BM Hamiltonian then takes the same block form as Eq.~\eqref{eq:BM_ham} with $h \to h'$, $g_i \to \bar{g}_i$ and $\bar g_1 = 0$, $\bar g_{2,3} = (M_t^{-1} - M_b^{-1})^T G_\pm$ where $G_\pm$ are primitive reciprocal lattice vectors of the undeformed lattice.

\subsection{QTM spectrum}
\label{app:BM_qtm}

The QTM spectrum as obtained from Eq.~\eqref{eq:QTM_spectrum} for the non-interacting BM model is shown in Fig.~\ref{fig:BM_qtm}. The unstrained spectrum matches the experimental results seen in Fig.~1 of Ref.~\cite{xiao2025interacting} slightly away from the magic angle, as well as the theoretical results in \cite{wei2025dirac} obtained using the exact equation for $\frac{d^2I}{dV_b^2}$, see Fig.~5 of that paper. The results for the strained case are in good agreement with Fig.~6 of Ref.~\cite{wei2025dirac}.

\begin{figure}
    \centering
\includegraphics[width=\columnwidth]{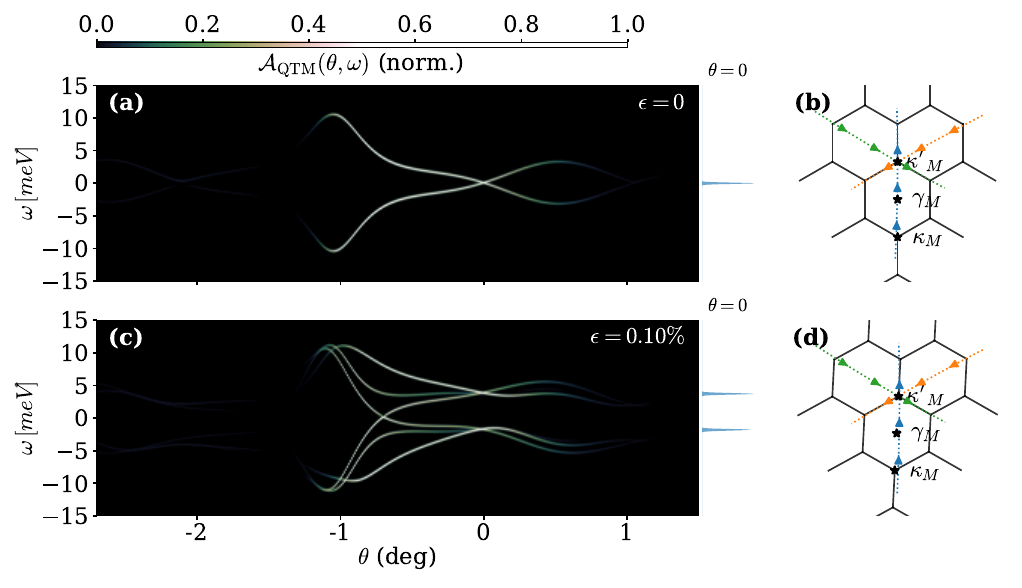}
\caption{QTM spectrum computed from the non-interacting BM model using Eq.~\eqref{eq:QTM_spectrum}. 
(a) Unstrained MATBG ($\epsilon=0$).
(b) Strained MATBG ($\epsilon = 0.1\%$, $\phi=15^\circ$). We use the parameters $w_0=88$ meV, $w_1=110$ meV, $v_D=658.2\text{ meV}\cdot\text{nm}$~\cite{wei2025dirac}.
\label{fig:BM_qtm}}
\end{figure}

\section{QTM spectrum: numerical details}
\label{app:qtm_numerics}

In this appendix we go through a number of key details regarding the numerical calculations for Fig.~\ref{fig:qtm}. The QTM spectrum is calculated using Eq.~\eqref{eq:QTM_spectrum}, which consists of two
factors: the spectral function $A^S(\mathbf{k},\omega)$ and the tunneling matrix element.
For the spectral function we follow the methods outlined in
Secs.~\ref{sec:controlled_expansion} and \ref{sec:strained}. The spectrum is calculated using the periodic THF formulation as defined
in Ref.~\cite{cualuguaru2023twisted, hu2025projected}.
For the matrix element, one must obtain
$\psi^\alpha_{tA/B}$, the weight of excitation $\alpha$ on the $A$/$B$ sublattice of
the top graphene layer. 

For a specific scanned $\mathbf k$, we obtain eigenvectors corresponding to different excitations. These are vectors of length $2+4\cdot N$, with $N$ the number of $c$ plane waves we use to write our periodic Hamiltonian, each corresponding to a plane wave at $\mathbf{k'+g}$, with $\mathbf g$ some reciprocal lattice vector and $\mathbf k'$ the momentum in the mini BZ. We take the $c$-orbital weight at the scanned momentum $\mathbf k=\mathbf{k'+g}$ itself, meaning the weight on the plane wave at the momentum scanned by the rotation of the tip $K_n$ point. The result is a vector $(f_1, f_2, c_1, c_2, c_3, c_4)$
in the THF orbital basis. Using this wave vector we extract the weights on the $A, B$ sublattices on the top MATBG layer using Ref.~\cite{cualuguaru2023twisted}. Note that the wavefunction coefficients in Ref.~\cite{cualuguaru2023twisted} are expressed in terms of the displacement
from the mBZ $K$-point,
\begin{equation}
    \mathbf{p} = \mathbf{k} - \eta\boldsymbol{\kappa}_M,
\end{equation}
where $\boldsymbol{\kappa}_M = \mathbf{q}_1$ is the mBZ $K$-point of valley $\eta$ that is aligned with the QTM tip for the QTM twist angle $\theta=0$, and
$\mathbf{k}$ is the momentum measured from $\gamma_M$~\cite{cualuguaru2023twisted}.

\subsection{$f$-electron sublattice spinors}

The $f$-electron creation operator is given by Eq.~(B30) of Ref.~\cite{cualuguaru2023twisted} as a Bloch
sum over plane waves from both graphene layers with momenta $\mathbf{k+Q}$, with $\mathbf Q \in \mathcal Q = \mathcal Q^{(+)}\oplus \mathcal Q^{(-)}$. The set $Q^{(\pm)}=\mathcal G\pm \mathbf q_1$ gives a honeycomb lattice of momenta in the top and bottom layer, respectively, with $\mathcal G$ the set of moir\'e reciprocal lattice vectors, and $\mathbf q_1$ the momentum of the $K$ point as measured from the $\gamma_M$ point (the specific $K$ point that is aligned with the tip for QTM angle $\theta_{QTM}=0$). The QTM matrix element
selects only the contribution from the plane wave at $\mathbf{k+Q}$ with $\mathbf{Q}=\eta\mathbf{q}_1$, corresponding to the $K$ point of the top layer, as detailed in Ref.~\cite{wei2025dirac}. Using the plane wave
coefficients $v^\eta_{\mathbf{Q},\beta,\alpha}(\mathbf{k})$ of Eqs.~(B20)--(B21) of
Ref.~\cite{cualuguaru2023twisted}, the resulting two-component sublattice spinors are
\begin{align}
    \bm{\psi}^{(f,1)} &= \frac{e^{+i\pi\eta/4}}{\sqrt{\Omega_M N_{f,\mathbf{k}}}}
    \begin{pmatrix}
        \alpha_1\sqrt{2\pi\lambda_1^2}\,e^{-|\mathbf{p}|^2\lambda_1^2/2} \\[4pt]
        \alpha_2\sqrt{2\pi\lambda_2^4}\,\eta(i\eta p_x - p_y)\,
        e^{-|\mathbf{p}|^2\lambda_2^2/2}
    \end{pmatrix},
    \label{eq:f1_spinor}
    \\[8pt]
    \bm{\psi}^{(f,2)} &= \frac{e^{-i\pi\eta/4}}{\sqrt{\Omega_M N_{f,\mathbf{k}}}}
    \begin{pmatrix}
        \alpha_2\sqrt{2\pi\lambda_2^4}\,\eta(-i\eta p_x - p_y)\,
        e^{-|\mathbf{p}|^2\lambda_2^2/2} \\[4pt]
        \alpha_1\sqrt{2\pi\lambda_1^2}\,e^{-|\mathbf{p}|^2\lambda_1^2/2}
    \end{pmatrix},
    \label{eq:f2_spinor}
\end{align}
where the rows correspond to sublattices $A$ and $B$ respectively, and the normalization
is
\begin{equation}
    \Omega_M N_{f,\mathbf{k}} = \sum_{\mathbf{p}'}
    \!\left[
        \alpha_1^2(2\pi\lambda_1^2)\,e^{-|\mathbf{p}'|^2\lambda_1^2}
        +\alpha_2^2(2\pi\lambda_2^2)\,|\mathbf{p}'|^2 e^{-|\mathbf{p}'|^2\lambda_2^2}
    \right],
    \label{eq:Nfk}
\end{equation}
with the sum running over $\mathbf{p}' = \mathbf{k}+\mathbf{Q}'$ for $\mathbf{Q}'\in\mathcal Q$. The numerical parameters such as $\alpha_1,\alpha_2, \lambda_1,\lambda_2$ are given in Table~\ref{table:wf_params}.

\subsection{$c$-electron sublattice spinors}

The $c$-electron creation operator is defined in Eq.~(B32) of Ref.~\cite{cualuguaru2023twisted}. The
plain wave coefficients $\tilde{u}^\eta_{\mathbf{Q},\beta,a}(\mathbf{k})$ are given at
$\mathbf{k}=0$ in Eqs.~(F1)--(F4) of Ref.~\cite{cualuguaru2023twisted}; for general $\mathbf{k}$ we apply the
small-$k$ approximation of Eq.~(F7), substituting $\mathbf{Q}\to\mathbf{Q}-\mathbf{k}$ in
those expressions. For the top layer this
gives $\mathbf{Q}-\mathbf{k}\to-\mathbf{p}$, so the spinors are evaluated using the
combinations $(-i\eta Q_x \pm Q_y)|_{\mathbf{Q}\to -\mathbf{p}}$. The four spinors
corresponding to the $\Gamma_3$ doublet ($a=1,2$) and $\Gamma_1\oplus\Gamma_2$ singlets
($a=3,4$) of Ref.~\cite{cualuguaru2023twisted} are
\begin{align}
    \bm{\psi}^{(c,1)} &= e^{-i\pi\eta/4}
    \begin{pmatrix}
        -\alpha_{c1}\sqrt{\dfrac{2\pi\lambda_{c1}^2}{\Omega_M N_{c1}}}\,
        e^{-|\mathbf{p}|^2\lambda_{c1}^2/2}
        \\[6pt]
        \alpha_{c2}\sqrt{\dfrac{\pi\lambda_{c2}^6}{\Omega_M N_{c1}}}\,
        (i\eta p_x+p_y)^2\,e^{-|\mathbf{p}|^2\lambda_{c2}^2/2}
    \end{pmatrix},
    \label{eq:c1_spinor}
    \\[8pt]
    \bm{\psi}^{(c,2)} &= e^{+i\pi\eta/4}
    \begin{pmatrix}
        \alpha_{c2}\sqrt{\dfrac{\pi\lambda_{c2}^6}{\Omega_M N_{c2}}}\,
        (i\eta p_x-p_y)^2\,e^{-|\mathbf{p}|^2\lambda_{c2}^2/2}
        \\[6pt]
        -\alpha_{c1}\sqrt{\dfrac{2\pi\lambda_{c1}^2}{\Omega_M N_{c2}}}\,
        e^{-|\mathbf{p}|^2\lambda_{c1}^2/2}
    \end{pmatrix},
    \label{eq:c2_spinor}
    \\[8pt]
    \bm{\psi}^{(c,3)} &= e^{-i\pi\eta/4}
    \begin{pmatrix}
        \alpha_{c3}\sqrt{\dfrac{2\pi\lambda_{c3}^4}{\Omega_M N_{c3}}}\,
        \eta(-i\eta p_x + p_y)\,e^{-|\mathbf{p}|^2\lambda_{c3}^2/2}
        \\[6pt]
        \alpha_{c4}\sqrt{\dfrac{\pi\lambda_{c4}^6}{\Omega_M N_{c3}}}\,
        (i\eta p_x-p_y)^2\,e^{-|\mathbf{p}|^2\lambda_{c4}^2/2}
    \end{pmatrix},
    \label{eq:c3_spinor}
    \\[8pt]
    \bm{\psi}^{(c,4)} &= e^{+i\pi\eta/4}
    \begin{pmatrix}
        \alpha_{c4}\sqrt{\dfrac{\pi\lambda_{c4}^6}{\Omega_M N_{c4}}}\,
        (i\eta p_x+p_y)^2\,e^{-|\mathbf{p}|^2\lambda_{c4}^2/2}
        \\[6pt]
        \alpha_{c3}\sqrt{\dfrac{2\pi\lambda_{c3}^4}{\Omega_M N_{c4}}}\,
        \eta(-i\eta p_x-p_y)\,e^{-|\mathbf{p}|^2\lambda_{c3}^2/2}
    \end{pmatrix}.
    \label{eq:c4_spinor}
\end{align}

\subsection{Matrix element calculation}

Given the $(f_1,f_2,c_1,c_2,c_3,c_4)$ eigenvector
$\mathbf{v}=(v_{f_1},v_{f_2},v_{c_1},v_{c_2},v_{c_3},v_{c_4})^T$ corresponding to some excitation, the
top-layer sublattice spinor used in Eq.~\eqref{eq:QTM_spectrum} is
\begin{equation}
    \begin{pmatrix}\psi^\alpha_{tA}\\\psi^\alpha_{tB}\end{pmatrix}
    =
    \sum_{b=1}^{2} v_{f_b}\,\bm{\psi}^{(f,b)}(\mathbf{p})
    +
    \sum_{a=1}^{4} v_{c_a}\,\bm{\psi}^{(c,a)}(\mathbf{p}).
    \label{eq:spinor_assembly}
\end{equation}
The QTM signal for that quasiparticle then contributes
$w_i \left| e^{i\eta\frac{2\pi(n-1)}{3}}\psi^\alpha_{tA}+\psi^\alpha_{tB}\right|^2$
to Eq.~\eqref{eq:QTM_spectrum}, with $\omega_i$ the residue of that excitation. The total QTM spectrum is a sum over all excitations.

The numerical parameters appearing in Eqs.~\eqref{eq:f1_spinor}--\eqref{eq:c4_spinor}
are listed in Table~\ref{table:wf_params}. The $c$ parameters are taken from
Eqs.~(F5)--(F6) of Ref.~\cite{cualuguaru2023twisted}. The $f$ parameters are taken for $\theta=1.04^\circ$ and $\omega_0/\omega_1=0.8$~\cite{cualuguaru2023twisted},
the closest available tabulated values to the $1.05^\circ$ twist angle used elsewhere in
this work.

\begin{table}[htbp]\centering
\begin{tabular}{c|c|c|c}
\multicolumn{4}{c}{$f$-electron parameters} \\
\hline
$\alpha_1$ & $\alpha_2$ & $\lambda_1\,[a_M]$ & $\lambda_2\,[a_M]$ \\
\hline
$0.818$ & $0.576$ & $0.179$ & $0.191$ \\
\hline\hline
\multicolumn{4}{c}{$c$-electron parameters ($\Gamma_3$ doublet, $a=1,2$)} \\
\hline
$\alpha_{c1}$ & $\alpha_{c2}$ & $\lambda_{c1}\,[a_M]$ & $\lambda_{c2}\,[a_M]$ \\
\hline
$0.3958$ & $0.9183$ & $0.2194$ & $0.3299$ \\
\hline
\multicolumn{2}{c|}{$N_{c1}=N_{c2}$} & \multicolumn{2}{c}{$1.2905$} \\
\hline\hline
\multicolumn{4}{c}{$c$-electron parameters ($\Gamma_1\oplus\Gamma_2$ singlets, $a=3,4$)} \\
\hline
$\alpha_{c3}$ & $\alpha_{c4}$ & $\lambda_{c3}\,[a_M]$ & $\lambda_{c4}\,[a_M]$ \\
\hline
$0.9257$ & $0.3783$ & $0.2430$ & $0.2241$ \\
\hline
\multicolumn{2}{c|}{$N_{c3}=N_{c4}$} & \multicolumn{2}{c}{$1.1102$} \\
\end{tabular}
\caption{Wavefunction parameters entering
Eqs.~\eqref{eq:f1_spinor}--\eqref{eq:spinor_assembly}. The $\lambda$ values are in units of
the moir\'e lattice constant $a_M$.}
\label{table:wf_params}
\end{table}

\end{widetext}

\end{document}